\documentclass[reprint,showpacs,preprintnumbers, amsmath,amssymb,aps,prc,]{revtex4-1}
\usepackage{amsfonts}
\usepackage{amsmath}
\usepackage{amssymb}
\usepackage{graphicx,color}
\usepackage{rotating}
\usepackage{dcolumn}% Align table columns on decimal 'point
\usepackage{bm,ulem,wasysym}% bold math
\usepackage[colorlinks=true, pdfstartview=FitV, linkcolor=blue, 
citecolor=blue,urlcolor=navyblue]{hyperref}

\begin{document}

\title{The symmetry energy $\gamma$ parameter of relativistic mean-field 
models} 

\author{Mariana Dutra} 
\affiliation{Departamento de Ci\^encias da Natureza, Universidade Federal Fluminense, 28895-532, 
Rio das Ostras, RJ, Brazil}
\author{Odilon Louren\c co}
\affiliation{Universidade Federal do Rio de Janeiro, 27930-560, Maca\'e, RJ, Brazil}
\author{Or Hen}
\affiliation{Massachusetts Institute of Technology, Cambridge, Massachusetts 02139, USA}
\author{Eliezer Piasetzky}
\affiliation{School of Physics and Astronomy, Tel Aviv University, Tel Aviv 69978, Israel}
\author{D\'ebora P. Menezes}
\affiliation{Depto de F\'isica - CFM - Universidade Federal de Santa Catarina, Florian\'opolis - SC 
- CP. 476 - CEP 88.040-900, Brazil}

\date{\today}

\begin{abstract}
The relativistic mean-field models tested in previous works against nuclear matter 
experimental values, critical parameters and macroscopic stellar properties are 
revisited and used in the evaluation of the symmetry energy $\gamma$ parameter obtained 
in three different ways. We have checked that independent of the choice made to 
calculate the $\gamma$ values, a trend of linear correlation is observed between $\gamma$ 
and the symmetry energy ($\mathcal{S}_0$) and a more clear linear relationship is 
established between $\gamma$ and the slope of the symmetry energy ($L_0$). These results 
directly contribute to the arising of other linear correlations between $\gamma$ and the 
neutron star radii of $R_{1.0}$ and $R_{1.4}$, in agreement with recent findings. 
Finally, we have found that short-range correlations induce two specific 
parametrizations, namely, \mbox{IU-FSU} and \mbox{DD-ME$\delta$}, simultaneously
compatible with the neutron star mass constraint of \mbox{$1.93\leqslant M_{\mbox{\tiny 
max}}/M_\odot\leqslant 2.05$} and with the overlap band for the $L_0\times\mathcal{S}_0$ 
region, to present $\gamma$ in the range of \mbox{$\gamma=0.25\pm0.05$}. 
\end{abstract}

\maketitle

\section{Introduction}
Since the introduction of the first models in nuclear physics, the main idea was to 
describe experimental data.  Not all nuclear models can be applied to the description of
nuclear matter but they are relevant nevertheless. Relativistic mean field (RMF) models
were developed to describe observables of nuclei, from which nuclear matter
parameters can be extracted. 

A detailed analysis of 263 RMF models based on pure neutron and symmetric nuclear matter 
properties was done in Ref.~\cite{PRC_055203} and only 35 of them were shown to satisfy 
important nuclear constraints. In a subsequent work, these models were used to analyse
stellar properties  related to largely studied astrophysical quantities, namely, neutron 
star masses and radii of the canonical neutron stars  (obtained from
observational data), the possible onset of the Direct Urca process and
sound velocity constraints~\cite{PRC_025806}. As a result, only 13
out of them produced neutron stars with 
maximum mass in the range of \mbox{$1.93\leqslant M_{\mbox{\tiny max}}/M_\odot\leqslant 
2.05$}~\cite{nature467-2010,science340-2013} as far as no hyperons
were considered, namely, one 
with density dependent couplings (\mbox{DD-F}) and one also incorporating 
scalar-isovector 
$\delta$ mesons (\mbox{DD-ME$\delta$}). The remaining parametrizations (BKA20, BKA22, 
BKA24, BSR8, BSR9, BSR10, BSR11, BSR12, FSUGZ03, \mbox{IU-FSU}, G2*) present constant 
couplings, nonlinear $\sigma$ and $\omega$ terms, and cross terms involving these fields. 
None of them could reproduce pulsars with $2 M_\odot$ if hyperons were included. More 
recently, the same models were revisited and their critical parameters were obtained 
\cite{PRCnew}. These critical parameters are the critical temperature, critical pressure 
and critical density, at which nuclear matter is no longer unstable and the liquid-gas 
phase transition ceases to exist~\cite{lg,vdw1,chomaz,vdw5,vdw2,vdw3,vdw6}. In this 
investigation, the models were divided into 6 categories (BKA, BSR, FSU, G2*, Z271 and 
DD). More experimental data is necessary, but so far, only two of them (Z271 and DD) 
provided critical temperatures close to existing ones. A clear correlation between the 
critical temperature and the compressibility was obtained. More details on these models 
are given along the paper.

In the same context, the symmetry energy~\cite{baldo} and its slope are very important 
quantities and in the last fifteen years, they were shown to be correlated with a series 
of physical properties, which we comment next. The symmetry energy is related to the 
nuclei neutron skin thickness, which in turn is related to neutron star radius: models 
that yield smaller neutron skin thickness in heavy nuclei, give rise to smaller neutron 
star radii~\cite{Piekarewicz_2001}. On the other hand, neutron skins are larger for 
models 
with higher slope  \cite{skin_2007}. Also, a strong correlation was observed 
between the neutron star radius and the variation of the slope at sub-threshold  
densities 
\cite{Lopes_2014}. The symmetry energy and the slope, however, can be easily controlled 
by 
the inclusion of  a $\omega-\rho$ \cite{skin_2007,Cavagnoli_2011,Prafulla_2012,EPJA_2014} 
or a $\sigma-\rho$ interaction~\cite{pais_2016} in non-linear
  models. The larger the value of the $\omega-\rho$ 
interaction, for instance, the lower the values of the symmetry energy and its slope. 
On the other hand, for density dependent models a change in the
density dependence of the $\rho$-meson coupling can modify the symmetry energy and its 
slope.

Besides the neutron skin thickness, the neutron star crust-core properties are also 
correlated with 
the slope of the symmetry energy, a fact that had already been observed in studies 
involving 
liquid-gas phase transitions, whose transition densities are approximately the same as the 
ones 
obtained as the separating densities from the pasta phase to homogeneous matter 
\cite{pasta_2009}.

We next reanalyze the parametrizations studied in Ref.~\cite{PRC_055203}, that we 
call {\it consistent relativistic mean field}~(CRMF) models from now on. In this context, 
the word ``consistent''  refers to those parametrizations that were
shown to satisfy the nuclear matter constraints in  ~Ref.~\cite{PRC_055203}. 
For these CRMF parametrizations, we evaluate the symmetry energy coefficient $\gamma$ in 
three 
different cases. Such a quantity is defined from 
$\mathcal{S}=\mathcal{S}^{kin}+\mathcal{S}^{pot}$, with 
$\mathcal{S}^{pot}(\rho)\propto(\rho/\rho_0)^\gamma$ and was first analyzed in Refs. 
\cite{Steiner_2009, Steiner_2010}.  Our aim is to look for possible correlations 
between the $\gamma$ parameter and some important nuclear matter and neutron star 
properties, namely the symmetry energy, its slope, and the radii of 1.0 and 1.4 solar 
mass 
neutron stars. Within the assumptions made in the present work, the $\gamma$ parameter 
fully defines the potential part of the symmetry energy and its density dependence. 
Theoretically, $\gamma$ is sensitive to the nucleon-nucleon interaction at very short 
distances and can be extracted from the existing calculations. We also investigate which 
parametrizations satisfy the ranges of $\gamma$ recently obtained in 
Refs.~\cite{gamma025,gamma072}. It is worth pointing out that from the experimental side, 
the $\gamma$ parameter is not directly measured but, as most of the bulk nuclear matter 
properties, it can be inferred from experiments. In order to be consistent with previous 
studies, the $\omega-\rho$ and $\sigma-\rho$ interactions that can be included in most 
models to {\it cure} their symmetry energy and slope values are left aside. They are just 
considered in the models that introduced them when they were proposed. 

This paper is organized as follows: the introduction is exhibited in Sec.~1. In 
Sec.~2, the formalism is introduced and three different forms of separate kinetic and 
potential parts of the symmetry energy are presented. In Sec.~3 the results are 
discussed. The summary is shown in Sec.~4.

\section{Formalism}

The analysis performed in Ref.~\cite{PRC_055203} pointed out to only 35 
parametrizations, out of 263 investigated, simultaneously approved in seven distinct 
nuclear matter constraints. These CRMF parametrizations had their bulk and 
thermodynamical 
quantities compared to respective theoretical/experimental data from symmetric nuclear 
matter (SNM), pure neutron matter (PNM), and a mixture of both, namely, symmetry energy 
and its slope evaluated at the saturation density, and the ratio of the symmetry energy 
at 
$\rho_0/2$ to its value at $\rho_0$ (MIX). These detailed constraints are specified in 
Table~\ref{constraints}.
\begin{table}[!htb]
\scriptsize
\begin{ruledtabular}
\caption{ \label{constraints}  Set of updated constraints (SET2a) used in 
Ref.~\cite{PRC_055203}. See that reference for more details.}
\begin{tabular}{lccc}
Constraint~ & Quantity      & Density Region       & Range \\  
SM1    & $K_0$     & at $\rho_0$  & 190 $-$ 270 MeV \\
SM3a   & $P(\rho)$ & $2<\frac{\rho}{\rho_0}<5$ & Band Region \\
SM4    & $P(\rho)$ & $1.2<\frac{\rho}{\rho_0}<2.2$    & Band Region\\
PNM1   & $\mathcal{E}_{\mbox{\tiny PNM}}/\rho$ & $0.017<\frac{\rho}{\rho_{\rm
o}}<0.108$   & Band Region \\
MIX1a  & $J$       & at $\rho_0$  & 25 $-$ 35 MeV \\
MIX2a  & $L_0$     & at $\rho_0$  & 25 $-$ 115 MeV \\
MIX4   & $\frac{\mathcal{S}(\rho_0/2)}{J}$  & at $\rho_0$ and $\rho_0/2$ & 0.57 $-$ 0.86\\
\end{tabular}
\end{ruledtabular}
\end{table}

In Table~\ref{models} we present a brief compilation of the structure and methods 
used in fitting the finite range RMF interactions in accordance with the macroscopic
constraints, and the data used for the fittings. For full explanation and details, we 
address the readers to the original papers and for a complete description of the 
relativistic mean-field theory, to Ref.~\cite{ppnp}.

Thirty out of the 35 parametrizations approved in matching the constraints analyzed in 
Ref.~\cite{PRC_055203} are of type~$4$, i.e., the Lagrangian density 
includes nonlinear $\sigma$ and $\omega$ terms and cross terms involving these fields. 
They are: BKA20, BKA22, BKA24, BSR8, BSR9, BSR10, BSR11, BSR12, BSR15, BSR16, 
BSR17, BSR18, BSR19, BSR20, \mbox{FSU-III}, \mbox{FSU-IV}, FSUGold, \mbox{FSUGold4}, 
FSUGZ03, FSUGZ06, G2*, \mbox{IU-FSU}, Z271s2, Z271s3, Z271s4, Z271s5, Z271s6, Z271v4, 
Z271v5, and Z271v6. They are described by the following Lagrangian density, 
\begin{eqnarray}
\mathcal{L}_{\mbox{\tiny NL}} &=& \overline{\psi}(i\gamma^\mu\partial_\mu - M)\psi 
+ g_\sigma\sigma\overline{\psi}\psi - g_\omega\overline{\psi}\gamma^\mu\omega_\mu\psi 
\nonumber \\
&-& \frac{g_\rho}{2}\overline{\psi}\gamma^\mu\vec{\rho}_\mu\vec{\tau}\psi
+ \frac{1}{2}(\partial^\mu \sigma \partial_\mu \sigma 
- m^2_\sigma\sigma^2) - \frac{A}{3}\sigma^3 
\nonumber\\
&-&  \frac{B}{4}\sigma^4 -\frac{1}{4}F^{\mu\nu}F_{\mu\nu} 
+ \frac{1}{2}m^2_\omega\omega_\mu\omega^\mu + 
\frac{C}{4}(g_\omega^2\omega_\mu\omega^\mu)^2 
\nonumber \\
&-& \frac{1}{4}\vec{B}^{\mu\nu}\vec{B}_{\mu\nu} + 
\frac{1}{2}m^2_\rho\vec{\rho}_\mu\vec{\rho}^\mu
+ \frac{1}{2}{\alpha_3'}g_\omega^2
g_\rho^2\omega_\mu\omega^\mu\vec{\rho}_\mu\vec{\rho}^\mu
\nonumber\\
&+& g_\sigma g_\omega^2\sigma\omega_\mu\omega^\mu
\left(\alpha_1+\frac{1}{2}{\alpha_1'}g_\sigma\sigma\right)
\nonumber\\
&+& g_\sigma g_\rho^2\sigma\vec{\rho}_\mu\vec{\rho}^\mu
\left(\alpha_2+\frac{1}{2}{\alpha_2'}g_\sigma\sigma\right),
\label{lomegarho}
\end{eqnarray}
with $F_{\mu\nu}=\partial_\nu\omega_\mu-\partial_\mu\omega_\nu$
and $\vec{B}_{\mu\nu}=\partial_\nu\vec{\rho}_\mu-\partial_\mu\vec{\rho}_\nu$. The nucleon 
rest mass is $M$ and the meson masses are $m_j$, for $j=\sigma,\omega,$ and $\rho$.

Other four CRMF approved parametrizations are density dependent (DD): 
\mbox{DD-F}, TW99, \mbox{DDH$\delta$} and \mbox{DD-ME$\delta$}. Their 
Lagrangian density reads:
\begin{eqnarray}
\mathcal{L}_{\mbox{\tiny DD}} &=& \overline{\psi}(i\gamma^\mu\partial_\mu - M)\psi 
+ \Gamma_\sigma(\rho)\sigma\overline{\psi}\psi 
- \Gamma_\omega(\rho)\overline{\psi}\gamma^\mu\omega_\mu\psi 
\nonumber\\
&-&\frac{\Gamma_\rho(\rho)}{2}\overline{\psi}\gamma^\mu\vec{\rho}_\mu\vec{\tau}
\psi + \Gamma_\delta(\rho)\overline{\psi}\vec{\delta}\vec{\tau}\psi 
- \frac{1}{4}F^{\mu\nu}F_{\mu\nu}
\nonumber \\
&+& \frac{1}{2}(\partial^\mu \sigma \partial_\mu \sigma - m^2_\sigma\sigma^2)
 + \frac{1}{2}m^2_\omega\omega_\mu\omega^\mu 
-\frac{1}{4}\vec{B}^{\mu\nu}\vec{B}_{\mu\nu}
\nonumber \\
&+& \frac{1}{2}m^2_\rho\vec{\rho}_\mu\vec{\rho}^\mu + 
\frac{1}{2}(\partial^\mu\vec{\delta}\partial_\mu\vec{\delta} 
- m^2_\delta\vec{\delta}^2),
\label{dldd}
\end{eqnarray}

\onecolumngrid
\textcolor{white}{asdfas}
\begin{table}[!htb]
\footnotesize
% \begin{ruledtabular}
\caption{\label{models} Structure of the RMF models and data used for fitting the 
finite range parametrizations  considered in the present work. NL: nonlinear model. DD: 
density dependent model. NAP: number of adjusted parameters. AT: additional terms in 
comparison with the standard nonlinear $\sigma^3-\sigma^4$ model with meson $\rho$ 
included.}
\begin{tabular}{llll}
Parametrization    & Type of model, n$^\circ$ of parameters, NAP, AT &  Data 
used for fitting purposes \\ \hline
BKA20, & NL, 12, 10 & Constraint properties of asymmetric nuclear 
matter for 26 different\\
BKA22,        & AT: $\omega^2$, $\sigma-\omega^2$, $\sigma^2-\omega^2$, $\sigma-\rho^2$  
                     & parametrizations:\\ 
BKA24~\cite{bka}     &                      & binding energies, charge radii for 
closed shell 
nuclei, \\
           &                      & neutron-skin thickness in the $^{\rm 208}$Pb nucleus:
                                    $0.20$, $0.22$, and $0.24$ fm.\\
\hline                                     
BSR8 to & NL, 14, 11   &  Binding energies: $^{\rm 16,24}$O, $^{\rm 
40,48}$Ca, $^{\rm 56,78}$Ni, $^{\rm 88}$Sr, $^{\rm 90}$Zr,$^{\rm 100,116,132}$Sn, \\ 
BSR12~\cite{bsr}   & AT: $\omega^2$, $\sigma-\omega^2$, $\sigma^2-\omega^2$, 
$\sigma-\rho^2$,                            &  and $^{\rm 208}$Pb nuclei,\\          
 
            & $\sigma^2-\rho^2$, $\omega^2-\rho^2$  &  charge radii: $^{\rm 16}$O, $^{\rm 
40,48}$Ca, $^{\rm 
                                            56}$Ni, $^{\rm 88}$Sr, $^{\rm 90}$Zr, $^{\rm 
116}$Sn, 
                                            and $^{\rm 208}$Pb nuclei  \\
            &                   &            neutron skin thickness: $^{\rm 208}$Pb\\
            &                   &          free parameters: 
                                           neutron-skin thickness $\Delta R= 0.16, 
                                           0.18,\cdots,0.28$ fm \\
            &                   &  and the $\omega$-meson self-coupling strength $\xi_0= 
0.03$\\
  
\hline         
BSR15 to & NL, 14, 11    &  The same as BSR8-BSR12 with $\xi_0 
= 0.06$ \\
BSR20~\cite{bsr}     & AT: $\omega^2$, $\sigma-\omega^2$, $\sigma^2-\omega^2$, 
$\sigma-\rho^2$, 
                          &  \\ 
& $\sigma^2-\rho^2$, $\omega^2-\rho^2$
\\ \hline
FSU-III,  & NL, 10,7  & Properties of asymmetric nuclear 
                                                            matter; the proton fraction 
in 
\\
FSU-IV\cite{PRC85-024302}  & AT: $\omega^2$, $\omega^2-\rho^2$       &  $\beta$-stable 
$npe\mu$ 
matter; the 
                                                             core-crust transition 
density 
and 
                                                             pressure \\
                  &   &  in neutron stars as predicted 
by 
                                                             FSUGold and IU-FSU;\\ 
                  &             &  free parameters: the coupling 
         
                                                             constants between the 
isovector 
                                                            $\rho$\\   
                  &            &  meson ($\Lambda_v$) and the 
isoscalar  
                                                       $\sigma$ and $\omega$ mesons 
($\Lambda_s$):\\
                  &                                       & FSU-III: $\Lambda_v = 0.00$ 
and 
                                                            $\Lambda_S = 0.02$\\  
                  &                                       & FSU-IV: $\Lambda_v = 0.00$ 
and 
                                                            $\Lambda_s = 0.04$\\ 
\hline                                                            
FSUGold~\cite{fsugold}    &  NL, 10, 8                  & Binding energies and 
charge radii 
                                                            magic nuclei for $^{\rm 
40}$Ca, $^{\rm 
                                                            90}$Zr, \\
                  &  AT: $\omega^2$, $\omega^2-\rho^2$  &  $^{\rm 116,132}$Sn, $^{\rm 
208}$Pb\\ 
\hline                   
FSUGold4~\cite{fsugold4}          &  NL, 10, 8, AT: $\omega^2$, $\omega^2-\rho^2$  & 
Adjusting the isovector parameters of 
                                                            the model $g_\rho$ and 
$\Lambda_v$ \\
\hline 
FSUGZ03,        &  NL, 14, 12                   & Binding energies: 
$^{\rm 16,24}$O, 
                                                            $^{\rm 40,48}$Ca, $^{\rm 
56,78}$ Ni, 
                                                            $^{\rm 88}$Sr, $^{\rm 90}$Zr, 
\\
FSUGZ06~\cite{fsugz}                 & AT: $\omega^2$, $\sigma-\omega^2$, 
$\sigma^2-\omega^2$, $\sigma-\rho^2$,    & $^{\rm 100,116,132}$Sn,$^{\rm 
208}$Pb\\  
                 &  $\sigma^2-\rho^2$, $\omega^2-\rho^2$                                  
 
   & charge rms radii:$^{\rm 
16}$O, 
                                                            $^{\rm 40,48}$Ca, $^{\rm 56}$ 
Ni, 
                                                            $^{\rm 88}$Sr, $^{\rm 90}$Zr, 
$^{\rm 
                                                            116}$Sn,$^{\rm 208}$Pb\\
                 &                                        & neutron-skin thickness 
for$^{\rm 
                                                            208}$Pb nucleus: $0.18 \pm 
0.01$ fm  \\  
                 &                                        & free parameters: $\zeta$ and 
$\xi$ 
                                                            corresponding to 
self-couplings for 
                                                            $\omega$\\ 
                 &                                        & and $\rho$ mesons: $\zeta = 
0.03, 0.06$ 
                                                            and $\xi = 0$ \\
\hline 
G2*~\cite{g2*}   &  NL, 12, 10               & Adjust the 
isovector-vector channel of 
                                                            the G2 parameter set. \\

                 & AT: $\omega^2$, $\sigma-\omega^2$, $\sigma^2-\omega^2$, 
$\sigma-\rho^2$  &\\
\hline 
IU-FSU~\cite{PRC82-055803}   &  NL, 10, 10     & Change the isoscalar 
parameter to $\xi 
                                                            = 0.03$; \\
                 & AT: $\omega^2$, $\omega^2-\rho^2$ & refitting of the isoscalar 
parameters 
to maintain the saturation \\
                 &                                        & properties of SNM of FSU;\\
                 &                                        & increase the 
isoscalar-isovector 
                                                         coupling constant to $\Lambda = 
0.046$\\  
\hline                               
Z271s2 to         & NL, 10, 8  & Model parameters 
used: Z271\\           
Z271s5~\cite{z271}                 & AT: $\omega^2$, $\sigma^2-\rho^2$    & free 
parameters: $\lambda_v = 
0$ and 
                                                            $\lambda_s = 
0.020,0.030,0.040,0.050$\\
\hline 
Z271v4 to        & NL, 10, 8   & The same as 
Z271s2-s5, but the\\         
Z271v6~\cite{z271}                 & AT: $\omega^2$, $\omega^2-\rho^2$   & free 
parameters are: $\lambda_s 
= 0$ and 
                                                            $\lambda_v = 
0.020,0.25,0.030$\\
\hline
DD-F~\cite{ddf}         & DD, 15, 12    & Properties of finite 
nuclei: binding 
                                                            energies, charge and\\
                 &                                     & diffraction radii, surface 
thicknesses, 
                                                         neutron skin \\
                 &                                     & in $^{\rm 208}$Pb, spin-orbit 
                                                         splittings\\
\hline                 
TW99~\cite{tw99}             & DD, 15, 12              & Fix the density 
dependence of the couplings from\\
                 &                                     & Dirac-Brueckner calculations of 
nuclear 
                                                         matter\\
                 &                                     & binding energies of symmetric 
nuclei ( 
                                                     $^{\rm 16}$O, $^{\rm 40}$Ca, $^{\rm 
56}$Ni) \\
                 &                                     & and neutron-rich nuclei ($^{\rm 
24}$O, 
                                                 $^{\rm 48}$Ca, $^{\rm 90}$Zr, $^{\rm 
208}$Pb).\\
\hline
DDH$\delta$~\cite{ddhd}      &  DD, 20, 16               & Reproduce bulk 
asymmetry parameter $a_4 = 
                                                         33.4$ MeV\\
                 &                   &  \\
\hline
DD-ME$\delta$~\cite{ddmed}    &  DD, 24, 14              & Finite nuclei 
and 
adjustment to ab initio 
                                                         calculations in infinite \\
                 &                                     &  nuclear matter\\
\hline                               
\end{tabular}
% \end{ruledtabular}
\end{table}
\twocolumngrid
%%%

where
\begin{eqnarray}
\Gamma_i(\rho) &=& \Gamma_i(\rho_0)f_i(x);\quad
f_i(x) = a_i\frac{1+b_i(x+d_i)^2}{1+c_i(x+e_i)^2},
\label{gamadefault}
\end{eqnarray}
for $i=\sigma,\omega$, and $x=\rho/\rho_0$. For the $\rho$ coupling one has
\begin{eqnarray}
\Gamma_\rho(\rho)=\Gamma_\rho(\rho_0)e^{-a_\rho(x-1)}.
\end{eqnarray}
The Lagrangian density describing the \mbox{DD-F} and TW99 parametrizations is the same 
as the one in Eq.~(\ref{dldd}) when the meson $\delta$ is not taken into account. For the 
\mbox{DD-ME$\delta$} parametrization, the couplings in Eq.~(\ref{gamadefault}) are valid 
for $i=\sigma,\omega,\rho$, and $\delta$. Finally, the \mbox{DDH$\delta$} model has the 
same coupling parameters as in Eq.~(\ref{gamadefault}) for the mesons $\sigma$ and 
$\omega$, but functions $f_i(x)$ given by 
\begin{eqnarray}
f_i(x)=a_ie^{-b_i(x-1)}-c_i(x-d_i),
\end{eqnarray}
for $i=\rho,\delta$.

Only one parametrization belongs to the nonlinear point coupling category, namely, the 
FA3~\cite{fa3}. In this kind of model, nucleons interact with each
other without explicitly including mesons~\cite{pc1,pc2,pc3,pc4}. Here, 
we do not investigate such model since in Ref~\cite{PRC_025806} we have shown it is not 
capable of generating a mass radius curve for neutron stars, due to a very particular 
behavior in the high-density regime, namely, a fall in the pressure versus energy density 
($\mathcal{E}$) curve near  $\mathcal{E} = 809$~MeV/fm$^3$. For that reason, we have 
decided to discard this particular parametrization.

All the details about the RMF approximation and related equations of state (EoS) are 
given in Ref.~\cite{PRC_055203} and will not be repeated here. Only the formulae 
necessary 
for the understanding of the present analysis are defined next.

The general definition of the symmetry energy reads as follows,
\begin{eqnarray}
\mathcal{S}(\rho) &=& \frac{1}{8}\frac{\partial^2(\mathcal{E}/\rho)}{\partial
y^2}\bigg|_{\rho,y=1/2} = 
\mathcal{S}^{kin}(\rho) + \mathcal{S}^{pot}(\rho),
\end{eqnarray}
where $y=\rho_p/\rho$ is the proton fraction of the system with $\rho_p$ being the proton 
density. By using such an expression, we compute the kinetic and potential contributions 
of the symmetry energy slope
\begin{eqnarray}
L(\rho)&=&3\rho\frac{\partial\mathcal{S}}{\partial\rho}
= 3\rho\frac{\partial\mathcal{S}^{kin}}{\partial\rho} +
3\rho\frac{\partial\mathcal{S}^{pot}}{\partial\rho}\nonumber\\
&=& L^{kin}(\rho) + L^{pot}(\rho).
\label{gen}
\end{eqnarray}

If we consider the potential part of the symmetry energy written as a power-law in 
density according to
\begin{eqnarray}
\mathcal{S}^{pot}(\rho) =
  \mathcal{S}^{pot}_0(\rho/\rho_0)^\gamma\equiv S^{pot}_{\mbox{\tiny approx.}}(\rho),
\label{powerlaw}
\end{eqnarray}
it is possible to express $L_0$ as
\begin{equation}
L_0=
3\rho_0 \left[ \left(\frac{\partial\mathcal{S}^{kin}}{\partial\rho}\right)_{\rho=\rho_0}
+ \frac{\gamma}{\rho_0} \mathcal{S}^{pot}_0 \right].
\label{eli}
\end{equation}
By using Eq.~(\ref{gen}) at $\rho=\rho_0$ and comparing it to Eq.~(\ref{eli}), one can 
find $\gamma$ as in Ref.~\cite{gamma025}, namely,
\begin{equation}
\gamma=\frac{L_0-L_0^{kin}}{3\,\mathcal{S}^{pot}_0} =
\frac{L_0^{pot}}{3\,\mathcal{S}^{pot}_0},
\label{eqgamma}
\end{equation}
where $\mathcal{F}_0^{kin,pot}=\mathcal{F}^{kin,pot}(\rho_0)$, for 
$\mathcal{F}=\mathcal{S}$, $L$. In that reference~\cite{gamma025}, the authors also 
introduced effects from short-range correlations (SRC) between proton-neutron 
pairs~\cite{src1,src2,src3} in symmetric nuclear matter in order to provide an analytical 
expression for the kinetic part of the symmetry energy. From this expression, that we 
will 
also use in Sec.~2.3, they found the range of $-10\pm7.5$~MeV for the kinetic part of 
the symmetry energy at the saturation density, $\mathcal{S}^{kin}_0$, based on data from 
free proton-to-neutron ratios measured in intermediate energy nucleus-nucleus collisions. 
Such a range allowed the authors to predict the values of \mbox{$\gamma=0.25\pm0.05$}.

Another proposition for the calculation of the $\gamma$ value is given in 
Ref.~\cite{gamma072}, where no short-range correlations in the kinetic part of the 
symmetry energy is taken into account. In that case, the density dependence of the 
symmetry energy was given by
\begin{eqnarray}
\mathcal{S}(\rho) = \mathcal{S}^{kin}(\rho) + \mathcal{S}^{pot}(\rho) =
a(\rho/\rho_0)^{2/3} + b(\rho/\rho_0)^{\gamma},\quad
\end{eqnarray}
with $a=12$~MeV, $b=22$~MeV, and $\gamma$ possibly ranging from $0.5$ to $1.5$ 
corresponding respectively to a soft and a stiff dependence. In that 
paper~\cite{gamma072}, a constraint for the nuclear symmetry energy at suprasaturation 
densities was deduced from ASY-EOS experiment at GSI at twice saturation density, where 
the measurement of the elliptic flows of neutrons and light-charged particles in a 
gold-gold reaction resulted in \mbox{$\gamma = 0.72 \pm 0.19$}.

\subsection{Complete kinetic term (case 1)}

Here we considered the {\it complete} kinetic term for the different models. Within this 
assumption, the first term of the symmetry energy is the kinetic part and the remaining 
is treated as the potential part. For the kinetic part, the corresponding expressions for 
nonlinear and density dependent (with $\delta$ meson) RMF models are, 
\begin{eqnarray}
\mathcal{S}_i^{kin}(\rho) &=& \frac{k_F^2}{6{E_F^*}_i}
\label{skinc1}
\end{eqnarray}
where $i=$ NL, DD, with ${E_F^*}_i=(k_F^2+{M_i^*}^2)^{1/2}$ and 
\begin{eqnarray}
M^*_{\mbox{\tiny NL}} = M - g_\sigma\sigma,\quad M^*_{\mbox{\tiny DD}} = M - 
\Gamma_\sigma(\rho)\sigma,
\end{eqnarray}
for symmetric matter ($y=1/2$). The Fermi momentum is written in term of density as 
$k_F=(3\pi^2\rho/2)^{1/3}$.

The potential part of the symmetry energy is written as
\begin{eqnarray}
\mathcal{S}_{\mbox{\tiny NL}}^{pot}(\rho) &=& \frac{g_\rho^2}{8{m_\rho^*}^2}\rho, 
\label{spotc1nl} \\
\mathcal{S}_{\mbox{\tiny DD}}^{pot}(\rho) &=& \frac{\Gamma_\rho^2\rho}{8m_\rho^2} 
-\frac{(\Gamma_\delta/m_\delta)^2({M^*_{\mbox{\tiny 
DD}}})^2\rho}{2{E_F^{*2}}_{\mbox{\tiny DD}}\left[1+\left(\frac{\Gamma_\delta}{m_\delta} 
\right)^2A_{\mbox{\tiny DD}}\right]},
\label{spotc1dd}
\end{eqnarray}
where
\begin{eqnarray}
{m_\rho^*}^2&=&m_\rho^2 + g_\sigma g_\rho^2\sigma(2\alpha_2+\alpha_2' g_\sigma\sigma)
+\alpha_3'g_\omega^2g_\rho^2\omega_0^2,\,\mbox{and}\qquad\\
A_{\mbox{\tiny DD}} &=& \frac{2}{\pi^2}\int_0^{k_F}\frac{k^4dk}{[k^2+({M^*_{\mbox{\tiny 
DD}}})^2]^{3/2}}=3\left(\frac{\rho_s}{M^*_{\mbox{\tiny DD}}}
-\frac{\rho}{{E_F^*}_{\mbox{\tiny DD}}}\right).\nonumber\\
\end{eqnarray}
The mean-field value of the vector field $\omega_\mu$ is $\omega_0$, and $\rho_s$ is the 
scalar density.

The respective expressions for the different contributions of the symmetry energy slope, 
namely, $L^{kin}(\rho)$ and $L^{pot}(\rho)$, are obtained as indicated in 
Eq.~(\ref{gen}), for this case and the next ones. 

\subsection{``Free'' kinetic term (case 2)}

In this case we have separated the {\it really kinetic term}, the one without any 
dependence of the interaction with the mesons, from the rest of the symmetry energy. The 
expressions in this case read
\begin{eqnarray}
\mathcal{S}^{kin}_{\mbox{\tiny NL}}(\rho) = \mathcal{S}^{kin}_{\mbox{\tiny DD}}(\rho) = 
\frac{k_F^2}{6E_F},
\label{skinc2}
\end{eqnarray}
with $E_F=(k_F^2+M^2)^{1/2}$, for the kinetic part, and
\begin{eqnarray}
\mathcal{S}_{\mbox{\tiny NL}}^{pot}(\rho) &=& 
\frac{k_F^2}{6{E_F^*}_{\mbox{\tiny NL}}}-\frac{k_F^2}{6E_F} 
+ \frac{g_\rho^2}{8{m_\rho^*}^2}\rho, 
\label{spotc2nl}\\
\mathcal{S}_{\mbox{\tiny DD}}^{pot}(\rho) &=& 
\frac{k_F^2}{6{E_F^*}_{\mbox{\tiny DD}}} - \frac{k_F^2}{6E_F} 
+ \frac{\Gamma_\rho^2\rho}{8m_\rho^2}\nonumber
\\
&-&\frac{(\Gamma_\delta/m_\delta)^2({M^*_{\mbox{\tiny 
DD}}})^2\rho}{2{E_F^{*2}}_{\mbox{\tiny DD}}\left[1+\left(\frac{\Gamma_\delta}{m_\delta} 
\right)^2A_{\mbox{\tiny DD}}\right]},\quad\,\,\,\,\,
\label{spotc2dd}
\end{eqnarray}
for the potential one.

\subsection{Short range correlations (case 3)}
% \label{c3}

The idea here is to replace the kinetic part of the symmetry energy by that one proposed 
in 
Ref.~\cite{gamma025}, where the authors have considered $\mathcal{S}^{kin}$ as composed 
by a free gas model term added to a correction term $\Delta\mathcal{S}^{\rm kin}$ that 
takes into account short-range correlations between proton-neutron pairs in symmetric 
nuclear matter. Based on this procedure, we calculate the potential part of the symmetry 
energy as follows,
\begin{equation}
\mathcal{S}^{pot}_i(\rho) = \mathcal{S}_i(\rho) - \mathcal{S}^{\rm kin}_{\rm SRC}(\rho),
\label{eqesym}
\end{equation}
where $i=$ NL, DD. The expressions for the total symmetry energy $\mathcal{S}_i(\rho)$ are
given by the sum of Eqs.~(\ref{skinc1}) and~(\ref{spotc1nl}) for the NL model, or 
Eqs.~(\ref{skinc1}) and~(\ref{spotc1dd}) for the DD one, by using the formulae of case~1. 
Exactly the same expressions are found if the case~2 is taken into account, i. e., if the 
sum of Eqs.~(\ref{skinc2}) and~(\ref{spotc2nl}) is performed for the NL model, or the sum 
of Eqs.~(\ref{skinc2}) and~(\ref{spotc2dd}) is considered for the DD model. Finally, the 
kinetic part of the symmetry energy for the present case analysis, $\mathcal{S}^{\rm 
kin}_{\rm SRC}(\rho)$, is taken from Ref.~\cite{gamma025} as
\begin{eqnarray}
\mathcal{S}_{\rm SRC}^{\rm kin}(\rho) = \left(2^{2/3}-1\right)\frac{3k_F^2}{10M} - 
\Delta\mathcal{S}^{\rm kin}(\rho),
\label{skineli}
\end{eqnarray}
with
\begin{eqnarray}
\Delta\mathcal{S}^{\rm kin}(\rho) &=& 
\frac{c_0{k_F^0}^2}{2M\pi^2}\left[\lambda\left(\frac{\rho}{\rho_0}\right)^{1/3}
-\frac{8}{5}\left(\frac {\rho}{\rho_0}\right)^{2/3} 
+\frac{3\rho}{5\lambda\rho_0}\right],\nonumber\\
\label{src}
\end{eqnarray}
where the parameters $c_0=4.48$ and $\lambda=2.75$ are also taken from 
Ref.~\cite{gamma025}. 

\section{Results}

\begin{figure}[!htb]
\centering
\includegraphics[scale=0.35]{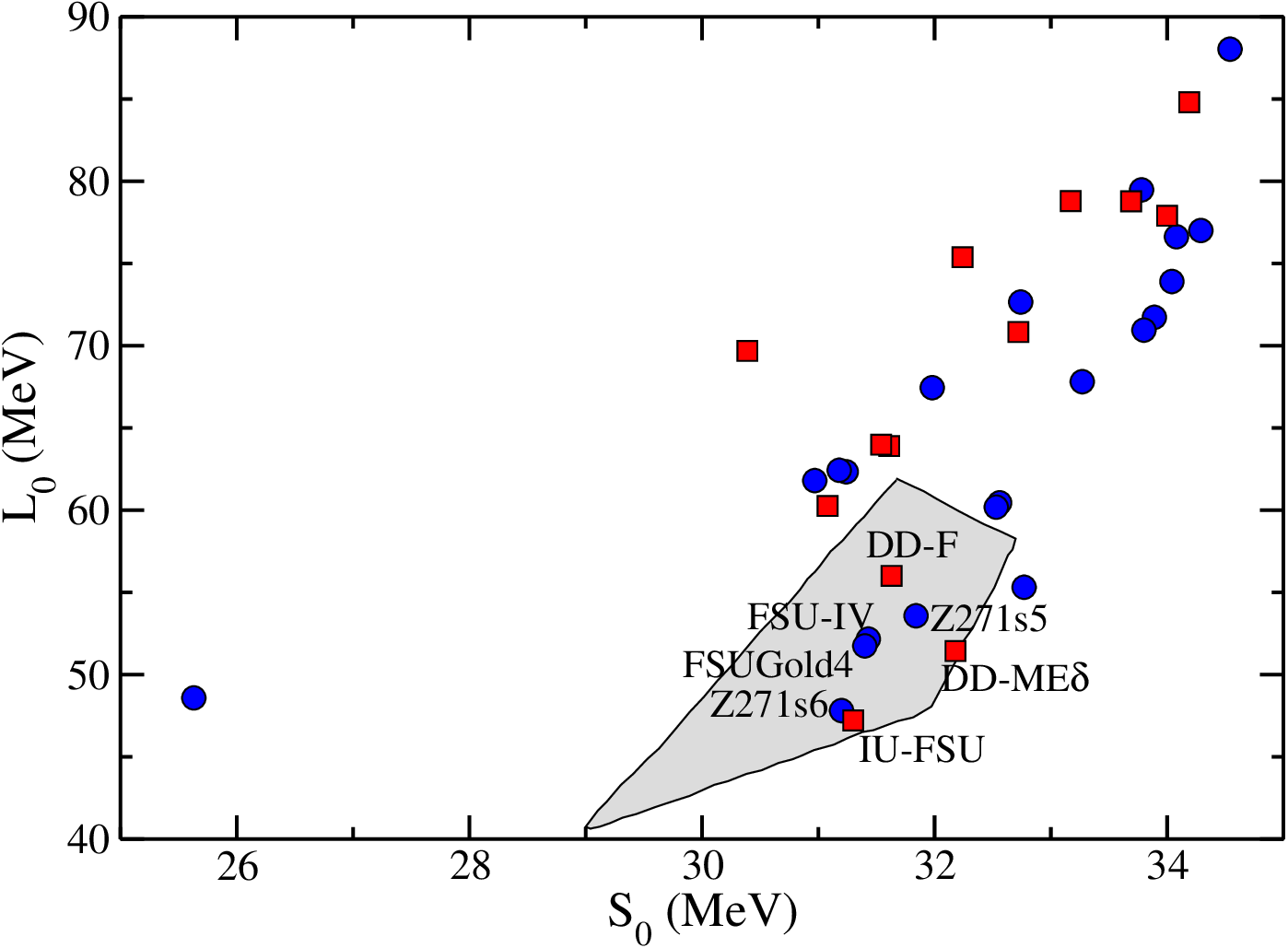}
\caption{Slope as a function of the symmetry energy for the CRMF models (all points). 
The 
gray band was extracted from Ref.~\cite{Lattimer_2013}. The squares represent 
parametrizations satisfying also the neutron star mass constraint 
of Refs.~\cite{nature467-2010,science340-2013}.}
\label{fig1}
\end{figure}

Let's start by revisiting the analysis of the correlation between the symmetry energy 
$\mathcal{S}_0=\mathcal{S}(\rho_0)$ and its slope $L_0=L(\rho_0)$, both at saturation 
density, whose data are shown respectively in columns 1 and 4 in Tables \ref{tablecase1}, 
\ref{tablecase2} and \ref{tablecase3} (that will be detailed latter) and are plotted in 
Fig.~\ref{fig1}, where squares refer to those parametrizations which also satisfy the 
macroscopic stellar properties of \mbox{$1.93\leqslant M_{\mbox{\tiny max}}/M_\odot 
\leqslant 2.05$} from Refs.~\cite{nature467-2010,science340-2013}. This correlation has 
already been 
extensively investigated, for instance, in 
Refs.~\cite{EPJA_2014,NScrust_book,Lattimer_2013,bianca} and only some of the points in 
Fig.~\ref{fig1} coincide with the overlap region of figure 2 in Ref.~\cite{Lattimer_2013} 
(gray band of our Fig.~\ref{fig1}). This means that the accepted range of values in 
Ref.~\cite{PRC_025806} is broader than the overlap of conditions shown in 
Ref.~\cite{Lattimer_2013}, namely, the overlap among constraints from nuclear masses, 
neutron skin thickness of Sn isotopes, dipole polarizability of $^{208}\rm Pb$, giant 
dipole resonances, isotope diffusion in heavy ion collisions, astrophysical observations, 
and neutron matter constraints.

We next obtain the $\gamma$ values by using Eq.~(\ref{eqgamma}) for some CRMF 
parametrizations and then also compare our values with the ranges proposed in 
Refs.~\cite{gamma025,gamma072}. In our analysis, we assume that the potential part of the 
symmetry energy can be written as in Eq.~(\ref{powerlaw}). Here, not all CRMF 
parametrizations are analyzed, but instead, only those in which the deviation defined 
by $\Delta(\rho)=|S^{pot}_{\mbox{\tiny model}}(\rho)-S^{pot}_{\mbox{\tiny 
approx.}}(\rho)|/S^{pot}_{\mbox{\tiny model}}(\rho)$ is less than a certain value, with 
$S^{pot}_{\mbox{\tiny approx.}}(\rho)=\mathcal{S}^{pot}_0(\rho/\rho_0)^\gamma$, see 
Eq.~(\ref{powerlaw}). As for each case one has different values for $\gamma$, the 
function 
$S^{pot}_{\mbox{\tiny approx.}}(\rho)$ exhibits different density dependences for the 
same 
parametrization. Therefore, each studied case produces different values of $\Delta(\rho)$ 
for the same parametrization. Since experimental values of $\gamma$ were extracted at 
suprasaturation density regime, we decided to investigate the values of $\Delta(\rho)$ at 
a range of $1\leqslant\rho/\rho_0\leqslant 4$, and define that only CRMF 
parametrizations, at this specific density range, presenting
$\Delta\leqslant 15\%$ are taken
into account in our study. By considering this analysis, we ensure a good 
agreement between the exact potential part of the symmetry energy, $S^{pot}_{\mbox{\tiny 
exact}}(\rho)$, of the CRMF parametrizations and the approximated form given in 
Eq.~(\ref{powerlaw}).

Our calculations were divided in the three different cases presented in Sec.~2. For each 
one, we construct a specific table where we use the symbol~$\checkmark$ to mark those 
models in which the constraint \mbox{$1.93\leqslant M_{\mbox{\tiny max}}/M_\odot\leqslant 
2.05$} for the maximum neutron star mass is also satisfied, the symbol~$\Box$ to mark 
those parametrizations presenting $\gamma$ parameter in the range of \mbox{$\gamma = 0.72 
\pm 0.19$}, and the symbol~$\boxtimes$ to identify parametrizations for which 
\mbox{$\gamma=0.25\pm 0.05$}. 

We next make a distinction between the two first cases (1 and 2) and case 3. For 
the last one presented in Sec.~3.3, we have considered short-range correlations in the 
kinetic part of the symmetry energy. For cases~1 and 2 such an effect is not taken into 
account. For this reason, we compare in next subsections the obtained values for the 
$\gamma$ parameter in cases~1 and~2 only with the experimental range of 
\mbox{$\gamma=0.72\pm0.19$}, since such range was obtained without SRC effects, according 
to Ref.~\cite{gamma072}. Furthermore, we name the parameters 
calculated from cases 1 and~2 respectively, as $\gamma_1$ and $\gamma_2$.

For case~3, the $\gamma$ parameter obtained with SRC effects is named
$\gamma_3$. It is compared only with the experimental range of
\mbox{$\gamma=0.25\pm 0.05$}
because this range was proposed in Ref.~\cite{gamma025} with SRC included in the analysis.

\subsection{Case 1}
\begin{figure}[!htb]
\centering
\includegraphics[scale=0.35]{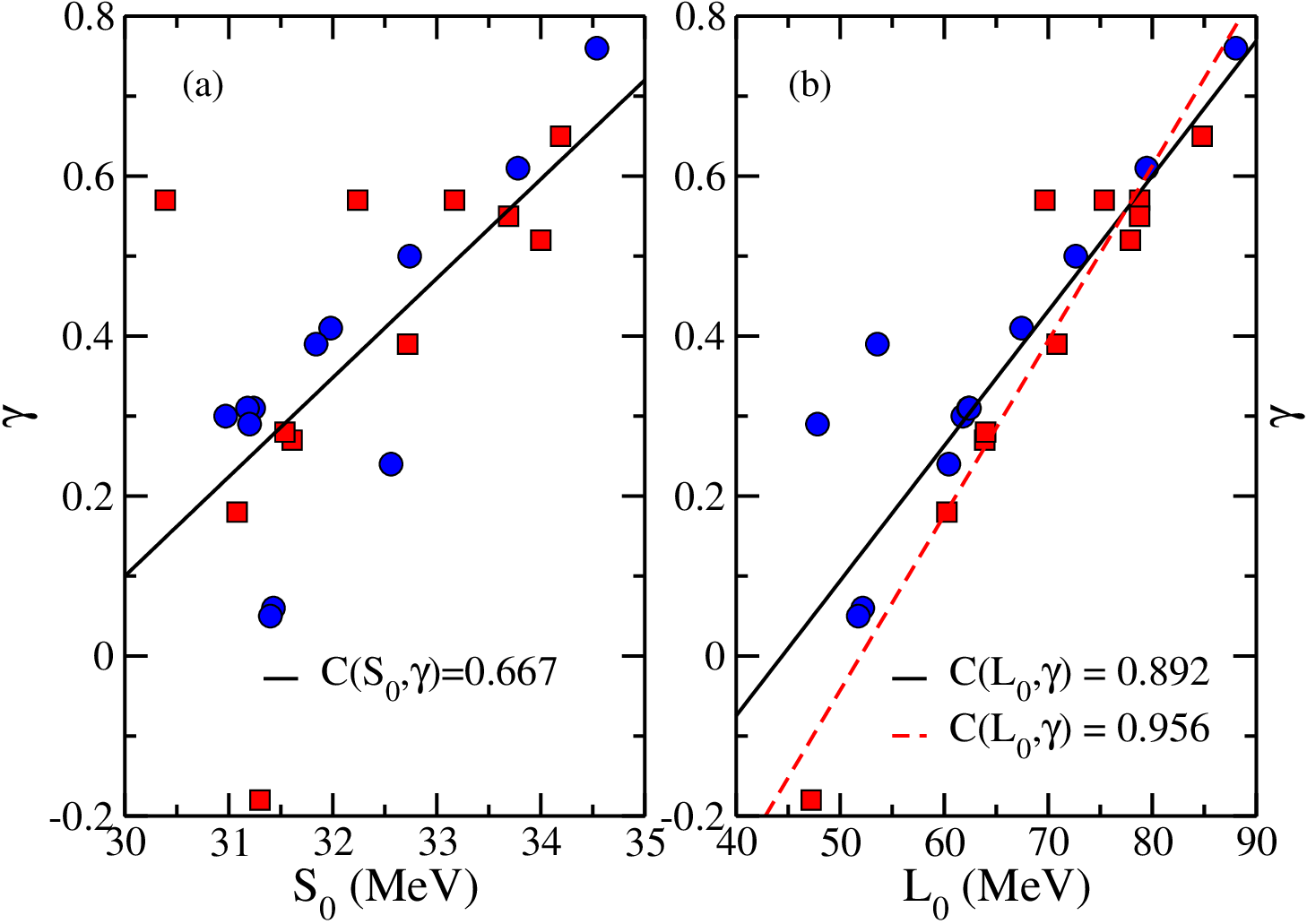}
\caption{$\gamma_1$ as a function of (a)~symmetry energy, and (b)~its slope, both at 
$\rho=\rho_0$, for the models displayed in Table~\ref{tablecase1} (all points). Squares 
represent 
parametrizations also satisfying the neutron star mass constraint of \mbox{$1.93\leqslant 
M_{\mbox{\tiny max}}/M_\odot\leqslant
2.05$}~\cite{nature467-2010,science340-2013}. The solid 
and dashed lines are fitting curves.}
\label{fig_case1}
\end{figure}

The $\gamma$ values calculated from case~1 ($\gamma_1$) are presented in 
Table~\ref{tablecase1}. From this table we notice only 7 parametrizations with $\gamma_1$ 
in the range of \mbox{$\gamma=0.72\pm0.19$}. Furthermore, with data taken from columns 1, 
4 and 7, we investigate possible correlations between $\gamma_1$ and the isovector 
quantities at the saturation density. The results are depicted in Fig.~\ref{fig_case1}.

% \newpage
\begin{table}[!htb]
\scriptsize
\begin{ruledtabular}
\caption{ \label{tablecase1}  Symmetry energy and its slope, with the respective 
kinetic and potential parts, all of them at $\rho=\rho_0$, obtained from the case~1 
analysis for the CRMF parametrizations presenting $\Delta(\rho)\leqslant 15\%$ at a 
density range of $1\leqslant\rho/\rho_0\leqslant 4$ (see main text). The meaning of 
symbols 
$\checkmark$ and $\Box$ is also defined in the text.}
\scriptsize
\begin{tabular}{lcccccccc}
 Models  & $\mathcal{S}_0$ & $\mathcal{S}_0^{kin}$ & $\mathcal{S}_0^{pot}$ & 
$L_0$ 
& $L_0^{kin}$  & $L_0^{pot}$ & $\gamma_1$  \\
& (MeV) &   (MeV)   &   (MeV)   &   (MeV)  &   (MeV)  &   (MeV) & \\
\hline
BKA20 $\checkmark$ $\Box$ & $32.24$ & $16.58$ & $15.66$ & $75.38$ & $48.47$ & $26.91$ & 
$0.57$\\
BKA22 $\checkmark$ $\Box$ & $33.17$ & $17.44$ & $15.73$ & $78.79$ & $52.12$ & $26.67$ & 
$0.57$ \\
BKA24 $\checkmark$ $\Box$ & $34.19$ & $17.54$ & $16.65$ & $84.80$ & $52.09$ & $32.70$ & 
$0.65$ \\
\hline
BSR8 $\checkmark$ & $31.08$ & $17.47$ & $13.61$ & $60.25$ & $52.78$ & $7.47$ & $0.18$ \\
BSR9 $\checkmark$ & $31.61$ & $17.57$ & $14.05$ & $63.89$ & $52.41$ & $11.49$ & $0.27$ \\
BSR10 $\checkmark$ & $32.72$ & $17.63$ & $15.09$ & $70.83$ & $53.09$ & $17.74$ & $0.39$ \\
BSR11 $\checkmark$ $\Box$ & $33.69$ & $17.47$ & $16.22$ & $78.78$ & $51.89$ & $26.89$ 
& $0.55$ \\
BSR12 $\checkmark$ & $34.00$ & $17.47$ & $16.53$ & $77.90$ & $52.30$ & $25.60$ & $0.52$ \\
BSR15 & $30.97$ & $17.33$ & $13.65$ & $61.79$ & $49.34$ & $12.45$ & $0.30$ \\
BSR16 & $31.24$ & $17.35$ & $13.90$ & $62.33$ & $49.41$ & $12.92$ & $0.31$ \\
BSR17 & $31.98$ & $17.38$ & $14.60$ & $67.44$ & $49.50$ & $17.93$ & $0.41$ \\
BSR18 & $32.74$ & $17.39$ & $15.35$ & $72.65$ & $49.48$ & $23.17$ & $0.50$ \\
BSR19 $\Box$ & $33.78$ & $17.40$ & $16.38$ & $79.47$ & $49.52$ & $29.96$ & $0.61$ \\
BSR20 $\Box$ & $34.54$ & $17.40$ & $17.14$ & $88.03$ & $49.10$ & $38.93$ & $0.76$ \\
\hline
FSU-IV & $31.43$ & $17.45$ & $13.98$ & $52.16$ & $49.72$ & $2.44$ & $0.06$ \\
FSUGold & $32.56$ & $17.45$ & $15.11$ & $60.44$ & $49.72$ & $10.72$ & $0.24$ \\
FSUGold4 & $31.40$ & $17.37$ & $14.03$ & $51.74$ & $49.43$ & $2.31$ & $0.05$ \\
FSUGZ03 $\checkmark$ & $31.54$ & $17.57$ & $13.98$ & $63.98$ & $52.40$ & $11.58$ &$0.28$\\
FSUGZ06 & $31.18$ & $17.35$ & $13.83$ & $62.42$ & $49.42$ & $13.00$ & $0.31$ \\
IU-FSU $\checkmark$ & $31.30$ & $17.94$ & $13.36$ & $47.21$ & $54.42$ & 
\hspace{-0.25cm}$-7.21$ & \hspace{-0.25cm}$-0.18$ \\
\hline
G2* $\checkmark$ $\Box$ & $30.39$ & $16.61$ & $13.77$ & $69.68$ & $46.31$ & $23.37$ 
& $0.57$\\
\hline
Z271s5 & $31.84$ & $13.82$ & $18.02$ & $53.57$ & $32.55$ & $21.02$ & $0.39$ \\
Z271s6 & $31.20$ & $13.82$ & $17.38$ & $47.81$ & $32.55$ & $15.25$ & $0.29$ \\
\end{tabular}
\end{ruledtabular}
\end{table}

From Fig.~\ref{fig_case1}{(a)}, we observe a trend of linear correlation 
between $\gamma_1$ and $\mathcal{S}_0$. A quantitative measurement of such finding can be 
given by calculation of the Pearson's correlation coefficient, defined as in 
Ref.~\cite{brandt}. Two different quantities $A$ and $B$ are strongly correlated within a 
linear relationship the more the coefficient correlation $C(A,B)$ is near to $1$, or $-1$ 
in the case of a negative linear dependence. In the case of the $\mathcal{S}_0$ 
dependence of $\gamma_1$, we found $C(\mathcal{S}_0,\gamma_1)=0.667$.

By performing the same study in the $\gamma_1\times L_0$ data, we noticed a better linear 
correlation than in the previous case, since the correlation coefficient resulted in 
$C(L_0,\gamma_1)=0.892$, see Fig.~\ref{fig_case1}{(b)}. Moreover, if we 
restrict our analysis only to the points corresponding to the models in which the neutron 
star mass constraint is satisfied (squares), we see that the linear correlation is still 
stronger in comparison to the one exhibited with all points. The correlation coefficient 
for the square points in Fig.~\ref{fig_case1}{(b)} is $C(L_0,\gamma_1)=0.956$.

\subsection{Case 2}

The $\gamma_2$ values calculated here are presented in Table~\ref{tablecase2}. 
Since in this case we have prevented the kinetic part of the symmetry energy from any 
influence of the effective mass, and as a consequence of the scalar meson effects, 
one can see here that \mbox{$\mathcal{S}_0^{kin}={k_F^0}^2/(6E_F^0)$} differs from each 
parametrization only due to the Fermi energy \mbox{$E_F^0=({k_F^0}^2+M^2)^{1/2}$} at the 
saturation point. As \mbox{$k_F^0=(3\pi^2\rho_0/2)^{1/3}$}, and as for nuclear mean-field 
models the saturation density is well established closely around the value of 
$\rho_0=0.15$~fm$^{-3}$, it becomes clear that the parametrizations analyzed according to 
Eqs.~(\ref{skinc2})-(\ref{spotc2dd}) present values of $\mathcal{S}_0^{kin}$ in a very 
narrow band as one can see from Table~\ref{tablecase2}. For the same reason, the kinetic 
part of the symmetry energy slope, $L_0^{kin}$, is also constrained to a small range. 
Also from Table~\ref{tablecase2}, we see that for the case~2 analysis, a large number of 
parametrizations, namely, 20 of them, have $\gamma_2$ in the range of \mbox{$\gamma = 
0.72 
\pm 0.19$}. 

% \newpage
\begin{table}[!htb]
\scriptsize
\begin{ruledtabular}
\caption{ \label{tablecase2}  Symmetry energy and its slope, with the respective 
kinetic and potential parts, all of them at $\rho=\rho_0$, obtained from the case~2 
analysis for the CRMF parametrizations presenting $\Delta(\rho)\leqslant 15\%$ at a 
density range of $1\leqslant\rho/\rho_0\leqslant 4$ (see main text). The meaning of 
symbols 
$\checkmark$ and $\Box$ is also defined in the text.}
\begin{tabular}{lcccccccc}
 Models  & $\mathcal{S}_0$ & $\mathcal{S}_0^{kin}$ & $\mathcal{S}_0^{pot}$ & 
$L_0$ 
& $L_0^{kin}$ & $L_0^{pot}$ & $\gamma_2$  \\
&   (MeV) &   (MeV)   &   (MeV)   &   (MeV)  &   (MeV)  &   (MeV) & \\
\hline
BKA20 $\checkmark$ $\Box$ & $32.24$ & $11.15$ & $21.09$ & $75.38$ & $21.53$ & $53.85$ 
& $0.85$ \\
BKA22 $\checkmark$ $\Box$ & $33.17$ & $11.21$ & $21.96$ & $78.79$ & $21.64$ & $57.15$ 
& $0.87$ \\
BKA24 $\checkmark$ & $34.19$ & $11.20$ & $22.99$ & $84.80$ & $21.63$ & $63.17$ & $0.92$ \\
\hline
BSR8  $\checkmark$ $\Box$ & $31.08$ & $11.19$ & $19.88$ & $60.25$ & $21.62$ & $38.64$  
& $0.65$\\
BSR9  $\checkmark$ $\Box$ & $31.61$ & $11.21$ & $20.40$ & $63.89$ & $21.65$ & $42.24$  
& $0.69$\\
BSR10 $\checkmark$ $\Box$ & $32.72$ & $11.22$ & $21.50$ & $70.83$ & $21.66$ & $49.17$ 
& $0.76$ \\
BSR11 $\checkmark$ $\Box$ & $33.69$ & $11.19$ & $22.50$ & $78.78$ & $21.60$ & $57.18$ 
& $0.85$ \\
BSR12 $\checkmark$ $\Box$ & $34.00$ & $11.22$ & $22.78$ & $77.90$ & $21.66$ & $56.24$ 
& $0.82$ \\
BSR15 $\Box$ & $30.97$ & $11.13$ & $19.85$ & $61.79$ & $21.49$ & $40.30$ & $0.68$ \\
BSR16 $\Box$ & $31.24$ & $11.13$ & $20.11$ & $62.33$ & $21.50$ & $40.83$ & $0.68$ \\
BSR17 $\Box$ & $31.98$ & $11.17$ & $20.81$ & $67.44$ & $21.57$ & $45.87$ & $0.73$ \\
BSR18 $\Box$ & $32.74$ & $11.14$ & $21.59$ & $72.65$ & $21.52$ & $51.13$ & $0.79$ \\
BSR19 $\Box$ & $33.78$ & $11.19$ & $22.60$ & $79.47$ & $21.60$ & $57.87$ & $0.85$ \\
BSR20 & $34.54$ & $11.15$ & $23.38$ & $88.03$ & $21.54$ & $66.48$ & $0.95$ \\
\hline
FSU-III $\Box$ & $33.89$ & $11.26$ & $22.64$ & $71.72$ & $21.73$ & $49.99$ & $0.74$ \\
FSU-IV & $31.43$ & $11.26$ & $20.17$ & $52.16$ & $21.73$ & $30.43$  & $0.50$ \\
FSUGold4 & $31.40$ & $11.22$ & $20.18$ & $51.74$ & $21.66$ & $30.07$& $0.50$ \\
FSUGZ03 $\checkmark$ $\Box$ & $31.54$ & $11.21$ & $20.33$ & $63.98$ & $21.65$ & 
$42.33$ & $0.69$\\
FSUGZ06 $\Box$ & $31.18$ & $11.14$ & $20.04$ & $62.42$ & $21.51$ & $40.92$ & $0.68$ \\
\hline
G2* $\checkmark$ $\Box$ & $30.39$ & $11.52$ & $18.87$ & $69.68$ & $22.22$ & $47.46$ & 
$0.84$ \\
\hline
Z271s2 $\Box$ & $34.08$ & $11.27$ & $22.81$ & $76.62$ & $21.75$ & $54.87$ & $0.80$ \\
Z271s3 $\Box$ & $33.27$ & $11.27$ & $22.00$ & $67.81$ & $21.75$ & $46.05$ & $0.70$ \\
Z271s4 $\Box$ & $32.53$ & $11.27$ & $21.26$ & $60.18$ & $21.75$ & $38.43$ & $0.60$ \\
Z271s5 & $31.84$ & $11.27$ & $20.57$ & $53.57$ & $21.75$ & $31.82$ & $0.52$ \\
Z271s6 & $31.20$ & $11.27$ & $19.93$ & $47.81$ & $21.75$ & $26.05$ & $0.44$ \\
\hline
DD-ME$\delta$ $\checkmark$ & $32.18$ & $11.44$ & $20.74$ & $51.43$ & $22.08$ & $29.35$ & 
$0.47$ \\
\end{tabular}
\end{ruledtabular}
\end{table}

\begin{figure}[!htb]
\centering
\includegraphics[scale=0.33]{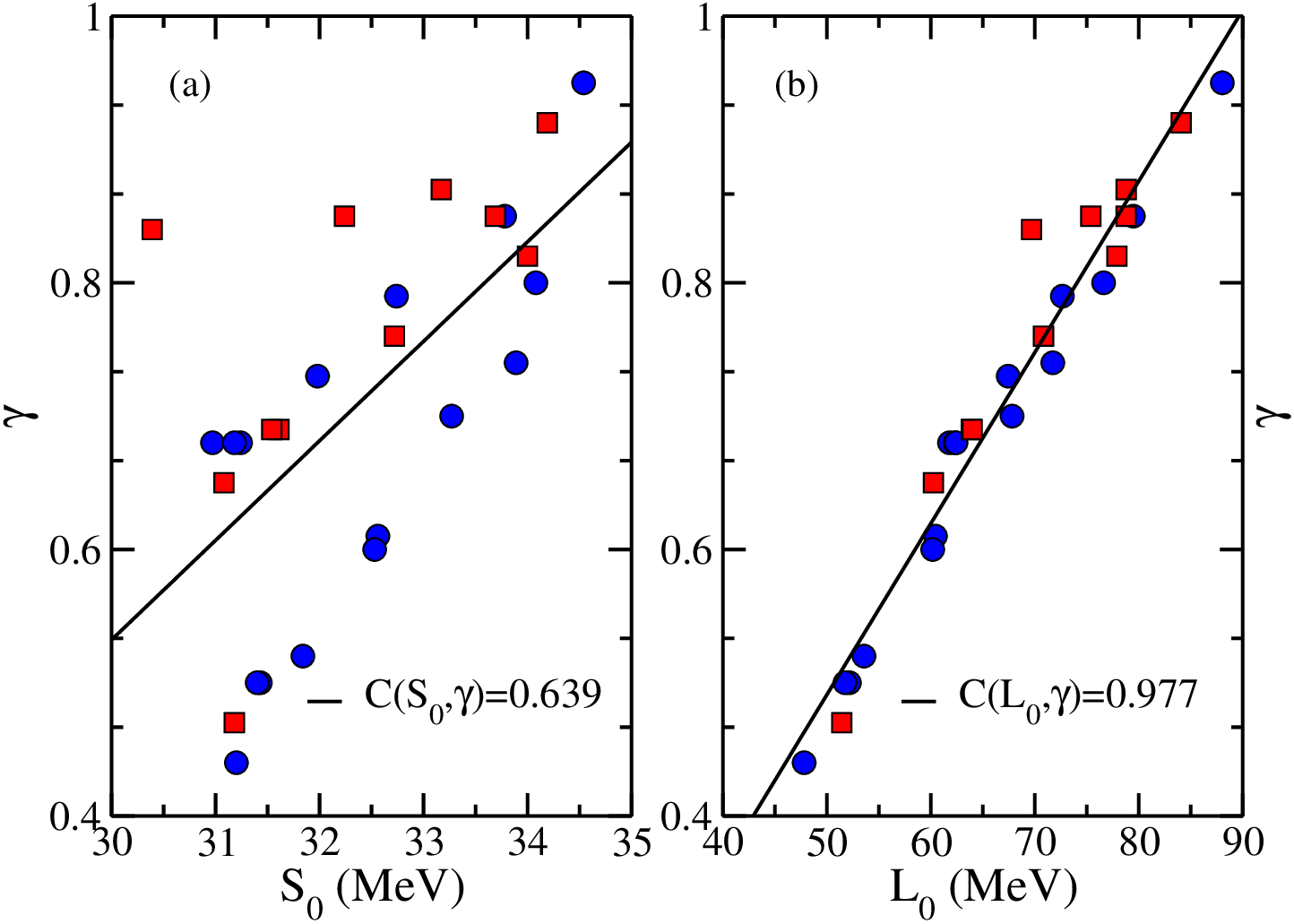}
\caption{$\gamma_2$ as a function of (a)~symmetry energy, and (b)~its slope, both at 
$\rho=\rho_0$, for the models of Table~\ref{tablecase2} (all points). Squares represent 
parametrizations also satisfying the neutron star mass constraint of \mbox{$1.93\leqslant 
M_{\mbox{\tiny max}}/M_\odot\leqslant 2.05$}~\cite{nature467-2010,science340-2013}. Solid 
lines: fitting curves.}
\label{fig_case2}
\end{figure}

As a further study, we analyse here the effect of the absence of the scalar interaction 
in the kinetic parts of the symmetry energy and its slope in the possible correlations of 
$\gamma_2$ with $\mathcal{S}_0$ and $L_0$. The results are shown in Fig.~\ref{fig_case2}.

From this figure one can see that the trend of linear correlation between $\gamma_2$ 
and $\mathcal{S}_0$ is worse when compared with case~1, since in case~2 one has 
$C(\mathcal{S}_0,\gamma_2)=0.639$. However, we see that the linear correlation 
$\gamma_2\times L_0$ is favored when the kinetic parts of the symmetry energy and its 
slope are free from the scalar interaction effects. The correlation coefficient in this 
case is $C(L_0,\gamma_2)=0.977$, a higher value than the corresponding one of the 
previous case, namely, $C(L_0,\gamma_1)=0.892$.

\subsection{Case 3}

The use of Eqs.~(\ref{eqesym})-(\ref{src}) along with Eq.~(\ref{gen}), all of them 
evaluated at $\rho=\rho_0$, allows the calculation of $\gamma_3$ from the definition 
given 
in Eq.~(\ref{eqgamma}). The results are presented in Table~\ref{tablecase3}. In our 
procedure, only the sum of the kinetic and potential parts of the symmetry energy 
matters. 
This sum does not change, and we extract the potential part by subtracting from the total 
(exact) value, the kinetic part with SRC included, as indicated in Eq.~(\ref{eqesym}). 

\begin{figure}[!htb]
\centering
\includegraphics[scale=0.33]{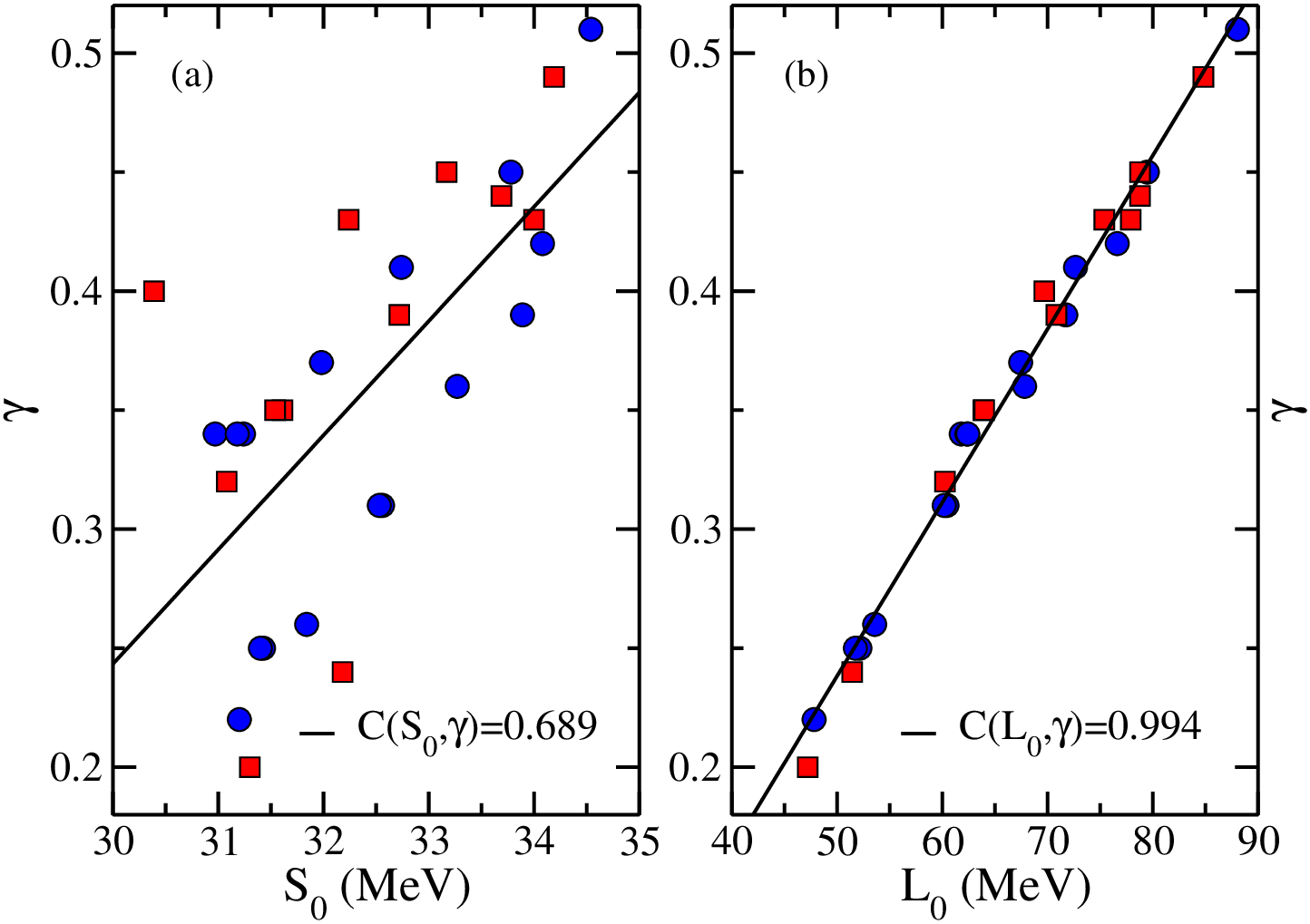}
\caption{$\gamma_3$ as a function of (a)~symmetry energy, and (b)~its slope, both at 
$\rho=\rho_0$, for the models of Table~\ref{tablecase3} (all points). Squares represent 
parametrizations also satisfying the neutron star mass constraint of \mbox{$1.93\leqslant 
M_{\mbox{\tiny max}}/M_\odot\leqslant 2.05$}~\cite{nature467-2010,science340-2013}. Solid 
lines: fitting curves.}
\label{fig_case3}
\end{figure}

% \newpage
\begin{table}[!htb]
\scriptsize
\begin{ruledtabular}
\caption{ \label{tablecase3}  Symmetry energy and its slope, with the respective 
kinetic and potential parts, all of them at $\rho=\rho_0$, obtained from the case~3 
analysis for the CRMF parametrizations presenting $\Delta(\rho)\leqslant 15\%$ at a 
density range of $1\leqslant\rho/\rho_0\leqslant 4$ (see main text). The meaning of 
symbols 
$\checkmark$ and $\boxtimes$ is also defined in the text.}
\begin{tabular}{lcccccccc}
 Models  & $\mathcal{S}_0$ & $\mathcal{S}_{\rm SRC,0}^{\rm kin}$ 
& $\mathcal{S}_0^{pot}$ & $L_0$ & $L_0^{kin}$ & $L_0^{pot}$ & $\gamma_3$  \\
&   (MeV) &   (MeV)   &   (MeV)   &   (MeV)  &   (MeV)  &   (MeV) & \\
\hline
BKA20 $\checkmark$ & $32.24 $ & $-9.31 $ & $41.55 $ & $75.38 $ & $21.21 $ & $54.16 $ & 
$0.43$ \\
BKA22 $\checkmark$ & $33.17 $ & $-9.36 $ & $42.53 $ & $78.79 $ & $21.33 $ & $57.46 $ & 
$0.45$ \\
BKA24 $\checkmark$ & $34.19 $ & $-9.35 $ & $43.54 $ & $84.80 $ & $21.31 $ & $63.48 $ & 
$0.49$ \\
\hline
BSR8  $\checkmark$ & $31.08 $ & $-9.35 $ & $40.43 $ & $60.25 $ & $21.30 $ & $38.95 $ & 
$0.32$ \\
BSR9  $\checkmark$ & $31.61 $ & $-9.37 $ & $40.98 $ & $63.89 $ & $21.34 $ & $42.55 $ & 
$0.35$ \\
BSR10 $\checkmark$ & $32.72 $ & $-9.37 $ & $42.09 $ & $70.83 $ & $21.35 $ & $49.48 $ & 
$0.39$ \\
BSR11 $\checkmark$ & $33.69 $ & $-9.34 $ & $43.03 $ & $78.78 $ & $21.29 $ & $57.49 $ & 
$0.44$ \\
BSR12 $\checkmark$ & $34.00 $ & $-9.37 $ & $43.37 $ & $77.90 $ & $21.35 $ & $56.55 $ & 
$0.43$ \\
BSR15 & $30.97 $ & $-9.29 $ & $40.26 $ & $61.79 $ & $21.17 $ & $40.62 $ & $0.34$ \\
BSR16 & $31.24 $ & $-9.30 $ & $40.54 $ & $62.33 $ & $21.18 $ & $41.15 $ & $0.34$ \\
BSR17 & $31.98 $ & $-9.33 $ & $41.31 $ & $67.44 $ & $21.25 $ & $46.18 $ & $0.37$ \\
BSR18 & $32.74 $ & $-9.31 $ & $42.04 $ & $72.65 $ & $21.20 $ & $51.45 $ & $0.41$ \\
BSR19 & $33.78 $ & $-9.34 $ & $43.13 $ & $79.47 $ & $21.29 $ & $58.19 $ & $0.45$ \\
BSR20 & $34.54 $ & $-9.31 $ & $43.85 $ & $88.03 $ & $21.22 $ & $66.80 $ & $0.51$ \\
\hline
FSU-III & $33.89 $ & $-9.40 $ & $43.30 $ & $71.72 $ & $21.43 $ & $50.30 $ & $0.39$ \\
FSU-IV $\boxtimes$ & $31.43 $ & $-9.40 $ & $40.83 $ & $52.16 $ & $21.43 $ & $30.73 $ 
&$0.25$\\
FSUGold & $32.56 $ & $-9.40 $ & $41.96 $ & $60.44 $ & $21.43 $ & $39.01 $ & $0.31$ \\
FSUGold4 $\boxtimes$ & $31.40 $ & $-9.37 $ & $40.77 $ & $51.74 $ & $21.35 $ & 
$30.38$&$0.25$\\
FSUGZ03 $\checkmark$ & $31.54 $ & $-9.37 $ & $40.91 $ & $63.98 $ & $21.34 $ & $42.64 $ & 
$0.35$ \\
FSUGZ06 & $31.18 $ & $-9.30 $ & $40.48 $ & $62.42 $ & $21.19 $ & $41.24 $ & $0.34$ \\
IU-FSU $\checkmark$ $\boxtimes$ & $31.30 $ & $-9.67 $ & $40.97 $ & $47.21 $ & $22.04 $ & 
$25.17 $ & $0.20$ \\
\hline
G2* $\checkmark$ & $30.39 $ & $-9.63 $ & $40.02 $ & $69.68 $ & $21.94 $ & $47.74 $ & 
$0.40$ \\
\hline
Z271s2 & $34.08 $ & $-9.41 $ & $43.49 $ & $76.62 $ & $21.45 $ & $55.18 $ & $0.42$ \\
Z271s3 & $33.27 $ & $-9.41 $ & $42.68 $ & $67.81 $ & $21.45 $ & $46.36 $ & $0.36$ \\
Z271s4 & $32.53 $ & $-9.41 $ & $41.94 $ & $60.18 $ & $21.45 $ & $38.74 $ & $0.31$ \\
Z271s5 $\boxtimes$ & $31.84 $ & $-9.41 $ & $41.25 $ & $53.57 $ & $21.45 $ & $32.12 $ 
&$0.26$\\
Z271s6 $\boxtimes$ & $31.20 $ & $-9.41 $ & $40.61 $ & $47.81 $ & $21.45 $ & $26.36 $ 
&$0.22$\\
\hline
DD-ME$\delta$ $\checkmark$ $\boxtimes$ & $32.18 $ & $-9.56 $ & $41.75 $ & $51.43 $ & 
$21.79 $ & $29.64 $ & $0.24$ \\ 
\end{tabular}
\end{ruledtabular}
\end{table}

From  Table~\ref{tablecase3} we see that $\mathcal{S}_{\rm SRC,0}^{\rm kin}$ has a 
negative value around $-9.3$~MeV for all parametrizations. Such a feature is a direct 
consequence of the short-range correlations between proton-neutron pairs in symmetric 
nuclear matter introduced in Ref.~\cite{gamma025}, that produced the expressions 
presented 
in Eqs.~(\ref{skineli})-(\ref{src}). Such a negative value for $\mathcal{S}_0^{kin}$ of 
the CRMF parametrizations is indeed consistent with the range of $-10\pm 7.5$~MeV found 
by the authors~\cite{gamma025} through the analysis of data from free proton-to-neutron 
ratios measured in intermediate energy nucleus-nucleus collisions.

We also see from Table~\ref{tablecase3} that the introduction of short-range correlations 
produces~6 parametrizations with $\gamma_3$ in the range of \mbox{$\gamma=0.25 \pm 
0.05$}. 

In Fig.~\ref{fig_case3}, we also investigate the $\gamma_3$ correlations. It is clear 
from this figure that the linear dependence between $\gamma_3$ and the isovector bulk 
parameters is still more favored when the short-range correlations are included in the 
CRMF parametrizations. The correlation coefficients obtained in this case are the closest 
to the unity, namely, $C(\mathcal{S}_0,\gamma_3)=0.689$ and $C(L_0,\gamma_3)=0.994$ in 
comparison with the respective quantities regarding the cases~1 and~2.

% \newpage

\subsection{Comments about the results}

Regarding the correlations found mainly between $\gamma$ and $L_0$, we remark that such 
result is not trivial, and the reason can be given from an analysis of 
Eq.~(\ref{eqgamma}). From such equation we can write:
\begin{eqnarray}
\gamma=\alpha L_0 + \beta,
\label{gammal0}
\end{eqnarray}
with $\alpha=1/(3S_0^{pot})$ and $\beta=-L_0^{kin}/(3S_0^{pot})$, i. e., a linear 
correlation between $\gamma$ and $L_0$ is obtained if $\alpha$ and $\beta$ are (ideally) 
constant numbers. In our study, we investigate whether $L_0^{kin}$ and $S_0^{pot}$ are 
close enough to constants for the sets of parametrization studied in each case. If this 
is 
the case, variations $\Delta\alpha$ and $\Delta\beta$ are close to zero. Since for 
each case studied we have the highest and lowest values of $S_0^{pot}$ and $L_0^{kin}$, 
it 
is possible to calculate $\Delta\alpha=\frac{1}{3{(S_0^{pot}})_{\mbox{\tiny high}}} - 
\frac{1}{3{(S_0^{pot}})_{\mbox{\tiny low}}}$, and 
$\Delta\beta=-\frac{{(L_0^{kin}})_{\mbox{\tiny high}}}{3{(S_0^{pot}})_{\mbox{\tiny 
high}}} + \frac{{(L_0^{kin}})_{\mbox{\tiny low}}}{3{(S_0^{pot}})_{\mbox{\tiny low}}}$. 
The absolute values of such calculations are shown in Table~\ref{tabcoef}, and as one can 
see, $\Delta\alpha$ and $\Delta\beta$ are decreasing quantities if we analyse such 
numbers 
from case~1 to~3. These behaviors explain the increasing coefficient correlation shown in 
Figs.~\ref{fig_case1} to~\ref{fig_case3}.

\begin{table}[!htb]
\scriptsize
\begin{ruledtabular}
\caption{ \label{tabcoef} Absolute values of $\Delta\alpha$ and $\Delta\beta$. 
Calculations for the three different cases analyzed.}
\begin{tabular}{ccc}
 case & $|\Delta\alpha|$ (MeV$^{-1}$) & $|\Delta\beta|$ \\
\hline
1 & $0.0064$ & $0.756$ \\
2 & $0.0034$ & $0.085$ \\
3 & $0.0007$ & $0.022$ \\ 
\end{tabular}
\end{ruledtabular}
\end{table}

This result is a consequence of the data previously presented in 
Tables~\ref{tablecase1}, \ref{tablecase2} and \ref{tablecase3}. From these tables, one 
can see that the values of $L_0^{kin}$ and $S_0^{pot}$ are closer to a constant value in 
case~3 than in case~2. Also, these values are closer to a certain constant value in 
case~2 
than in case~1.

One can see from such a table that the closer results are obtained for the cases in which 
the scalar attractive interaction is not taken into account in the kinetic part of the 
symmetry energy, i. e., cases~2 and~3, being the last one the case in which 
$\Delta\alpha$ 
and $\Delta\beta$ are closer to zero. 

\begin{figure}[!htb]
\centering
\includegraphics[scale=0.236]{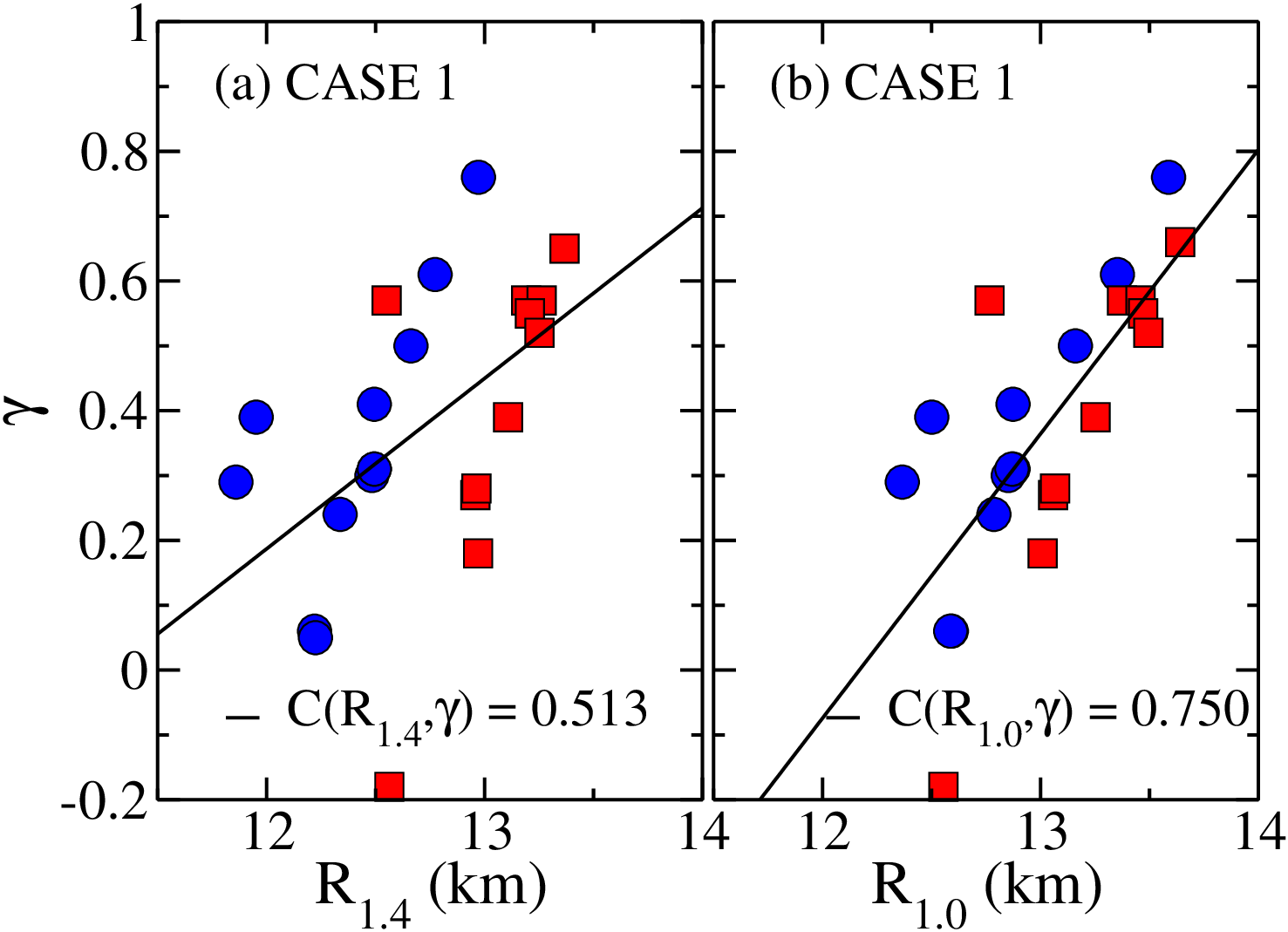}
\includegraphics[scale=0.236]{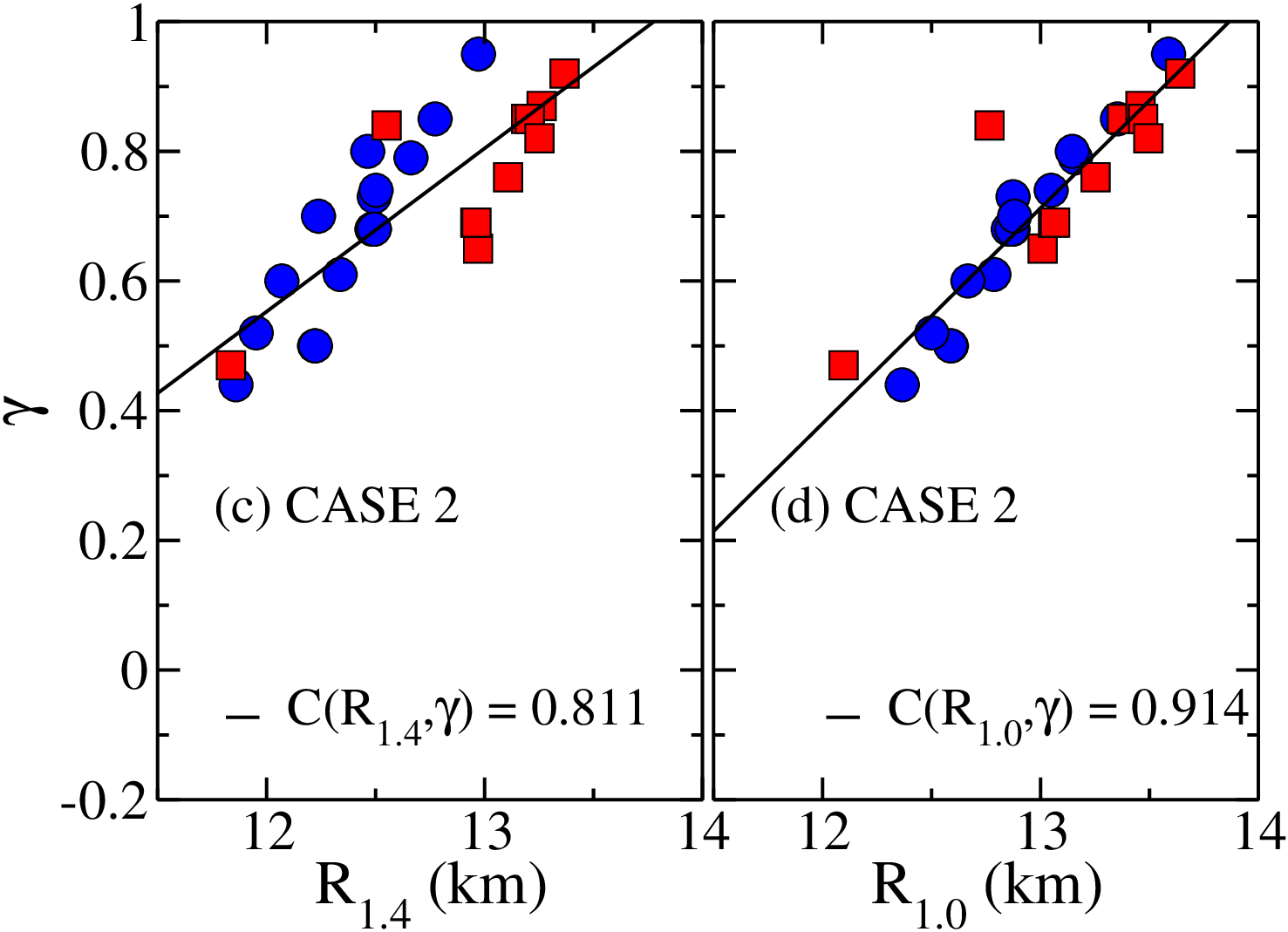}
\includegraphics[scale=0.236]{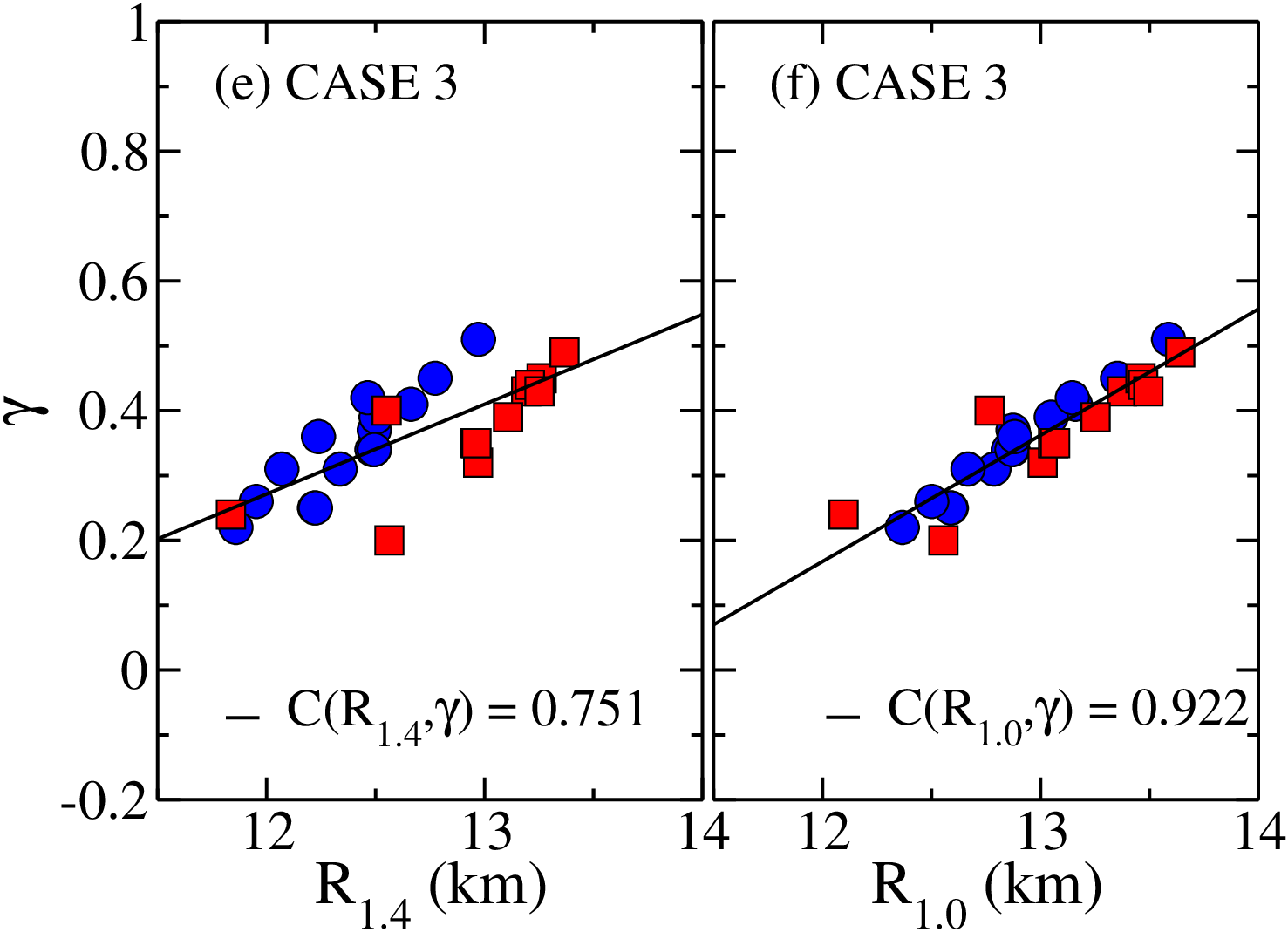}
\caption{$\gamma_1$, $\gamma_2$, and $\gamma_3$ as a function of the $R_{1.0}$ and 
$R_{1.4}$ neutron star radii for the CRMF parametrizations (all points). Squares 
represent 
parametrizations also satisfying the constraint of \mbox{$1.93\leqslant M_{\mbox{\tiny 
max}}/M_\odot\leqslant 2.05$}~\cite{nature467-2010,science340-2013}.}
\label{radii}
\end{figure}

As a direct application of this specific correlation, we also investigate whether 
$\gamma_1$, $\gamma_2$, and $\gamma_3$ obtained from the CRMF parametrizations in the 
three different cases studied and always calculated for symmetric
  nuclear matter, also correlates with neutron star radii. The motivation of 
such study comes from the results presented in Ref.~\cite{constanca}, in which authors 
found that for a class of 42 relativistic and Skyrme parametrizations, $L_0$ linearly 
depends on $R_{1.0}$ and $R_{1.4}$, namely, the radii of neutron stars presenting 
$M_{\mbox{\tiny star}}=M_\odot$ and~$1.4M_\odot$, respectively. The $R_{1.0}$ and
$R_{1.4}$ dependence of $\gamma_1$, $\gamma_2$, $\gamma_3$ related to those CRMF 
parametrizations analyzed here is shown in Fig.~\ref{radii}. In
  order to obtain stellar macroscopic properties, the same CRMF
  parametrizations are used, but now the models are subject to matter
  neutrality and $\beta$-equilibrium.

In order to generate the neutron star radii, we have joined the hadronic matter EoS from 
the CRMF parametrizations with those for electrons and muons. After that, the conditions 
of charge neutrality and chemical equilibrium were taken into account and the 
Baym-Pethick-Sutherland (BPS) equation of state~\cite{bps} for low densities was added to 
the EoS for hadrons and leptons. The resulting EoS was used as input to the 
Tolman-Oppenheimer-Volkoff equations~\cite{tov}. We address the reader to details 
regarding such calculations to Ref.~\cite{details}, for instance. 

From Fig.~\ref{radii}, we can conclude that the CRMF parametrizations also present a 
linear behavior concerning $\gamma_1$, $\gamma_2$, $\gamma_3$ and the neutron star radii. 
This result is entirely compatible with the findings of Ref.~\cite{constanca}. In that 
paper, a linear correlation between $L_0$ and the radii was found, and since in our study 
we have found a linear dependence for $\gamma_1$, $\gamma_2$, $\gamma_3$ and $L_0$, 
according to Eq.~(\ref{gammal0}), a direct consequence is the linear behavior described 
in 
Fig.~\ref{radii}. Also as in Ref.~\cite{constanca}, the correlations are stronger for the 
$R_{1.0}$ neutron star radius as the correlation coefficients $C(R_{1.0},\gamma)$ point 
out. Finally, as observed along all investigations, the linear dependence is intensified 
in 
cases~2 and~3 in which the effects of the scalar interaction are absent from the kinetic 
part of the symmetry energy.

\begin{figure}[!htb]
\centering
\includegraphics[scale=0.34]{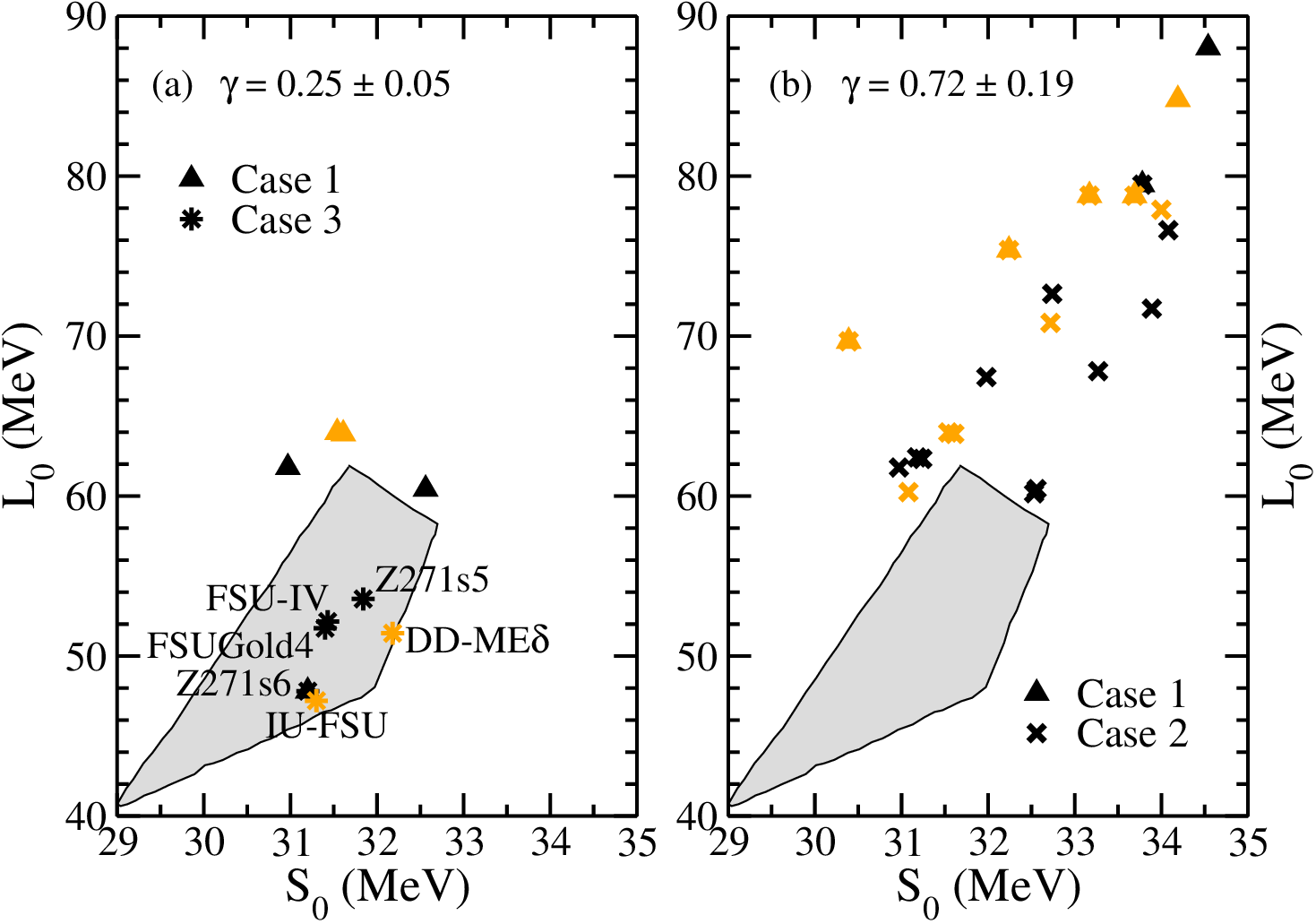}
\caption{Slope as a function of the symmetry energy for the models that produce the 
(a) lower and (b) higher $\gamma$ ranges discussed in the last section. The gray band was 
extracted from Ref.~\cite{Lattimer_2013}. Orange points represent those parametrizations 
in which the neutron star mass constraint of \mbox{$1.93\leqslant M_{\mbox{\tiny 
max}}/M_\odot\leqslant 2.05$}~\cite{nature467-2010,science340-2013} is verified.}
\label{fignew}
\end{figure}

Another point we remark here concerns the gray band of Fig.~\ref{fig1} for the CRMF 
models in the different cases studied. In order to do that, we start by redrawing such a 
figure for the different models that reproduce \mbox{$\gamma=0.25 \pm 0.05$}, panel (a), 
and \mbox{$\gamma=0.72 \pm 0.19$}, panel (b) in Fig.~\ref{fignew}.

It is worth noting that the CRMF parametrizations for which we
  have obtained the $\gamma$ 
parameters from case~3 are more compatible with the gray band proposed in 
Ref.~\cite{Lattimer_2013}, i. e., the short-range correlations effects induce the CRMF 
parametrizations to present the $\gamma$ parameter inside the range of 
\mbox{$\gamma=0.25\pm0.05$}, simultaneously being consistent with the overlap conditions 
of Ref.~\cite{Lattimer_2013} obtained from many experimental and observational data. For 
such a case, 6 parametrizations are inside the overlap band, namely, \mbox{IU-FSU}, 
\mbox{FSU-IV}, FSUGold4, Z271s5, Z271s6, and \mbox{DD-ME$\delta$}, with 2 of them, 
\mbox{IU-FSU} and \mbox{DD-ME$\delta$}, also satisfying the neutron star mass 
constraint of \mbox{$1.93\leqslant M_{\mbox{\tiny max}}/M_\odot\leqslant 
2.05$}~\cite{nature467-2010,science340-2013} and two of them, Z271s5 and Z271s6 yielding 
critical parameters close to the existing proposition of experimental values, according 
to the findings of Ref.~\cite{PRCnew}.

\section{Summary}

In summary, our calculations have shown that, independently of the choice made to obtain 
the $\gamma$ values (case~1, 2 or~3) for the CRMF models, a trend of linear correlation 
is observed between $\gamma_1$, $\gamma_2$, $\gamma_3$ and $\mathcal{S}_0$, and a more 
clear linear relationship is established regarding $\gamma_1$, $\gamma_2$, $\gamma_3$ and 
the slope of the symmetry energy at the saturation density, $L_0$. In cases~2 and~3, the 
last correlation is still more pronounced. Such effect arises due to the absence of the 
attractive interaction in the kinetic part of the symmetry energy. Furthermore, the 
short-range correlations introduced in the case~3 analysis intensify the linear $L_0$ 
dependence of $\gamma_3$ as seen in Fig.~\ref{fig_case3}{(b)}. These results can be used 
to determine other linear correlations of $\gamma_1$, $\gamma_2$, $\gamma_3$ and the 
neutron star radii of $R_{1.0}$ and $R_{1.4}$, as displayed in Fig.~\ref{radii}. Finally, 
specifically for case~3, two specific parametrizations, namely, \mbox{IU-FSU} and 
\mbox{DD-ME$\delta$} are shown to be compatible with the range of \mbox{$\gamma=0.25 \pm 
0.05$}~\cite{gamma025}, and simultaneously consistent with the neutron star mass 
constraint of Refs.~\cite{nature467-2010,science340-2013}, and other two, Z271s5 and 
Z271s6, simultaneously compatible with the range of \mbox{$\gamma=0.25 \pm 
0.05$}~\cite{gamma025} and with probable critical parameters experimental 
values~\cite{PRCnew}. The four parametrizations are consistent with the overlap band for 
the $L_0\times\mathcal{S}_0$ region described in Ref.~\cite{Lattimer_2013}, see 
Fig.~\ref{fignew}.

As a final remark, we remind the reader that we have only analyzed symmetric matter 
in this study for the calculation of the $\gamma$ values, but
the potential difference for neutrons and protons in neutron-rich 
matter and their density dependence, for instance, can also be calculated. However,
the results could not be compared with the existing $\gamma$
values. Furthermore, if we also want to investigate the 
momentum dependence in asymmetric matter, single particle 
potentials, which are different for neutrons and protons, have to be
taken into account, see Ref.~\cite{Bao_book,Bao_2004}. To obtain this
kind of dependence,  one would  need either a theory that uses
non-local interactions or a Thomas-Fermi calculation and 
both approaches are out of the scope of the present work.

\section*{Acknowledgments} 

This work is a part of the project INCT-FNA Proc. No. 464898/2014-5
and was partially supported by Conselho Nacional de
Desenvolvimento Cient\'ifico e Tecnol\'ogico (CNPq), Brazil under
grants 300602/2009-0 and 306786/2014-1. E.~P.  
acknowledges support from the Israel Science Foundation and thanks the Brazilian 
Physical Society and the organizers of the XXXVIII RTFNB (nuclear physics
annual  meeting) for the invitation in 2015, when this collaboration
started. O.~H. acknowledges the U.S. Department of Energy 
Office of Science, Office of nuclear physics program under award number DE-FG02-94ER40818.


\begin{thebibliography}{99}
\bibitem{PRC_055203} M. Dutra, O. Louren\c{c}o, S. S. Avancini, B. V. Carlson, A. 
Delfino, 
D. P. Menezes, C. Provid\^encia, S. Typel, and J. R. Stone,  Phys. Rev. C {\bf 90}: 
055203 (2014).

\bibitem{PRC_025806} M. Dutra, O. Louren\c co and D. P. Menezes, Phys. Rev. C {\bf 
93}: 025806 (2016); Phys. Rev. C {\bf 94}: 049901(E) (2016).

\bibitem{nature467-2010} P. B. Demorest, T. Pennucci, S. M. Ransom, M. S. E. Roberts, and 
J. W. T. Hessels, Nature {\bf 467}: 1081 (2010).

\bibitem{science340-2013} J. Antoniadis, P. C. C. Freire, N. Wex {\it et al.}: Science 
{\bf 340}: 448 (2013).

\bibitem{PRCnew} O. Louren\c co, M. Dutra and D. P. Menezes, Phys. Rev. C {\bf 95}: 
065212 (2017).
  
\bibitem{lg} G. Bertsch and P. J. Siemens, Phys. Lett. B {\bf 126}: 9
  (1983); J. Margueron and P. Chomaz, Phys. Rev. C {\bf 67}: 041602 (2003); 
C. Ducoin, Ph. Chomaz, and F. Gulminelli, Nucl. Phys. A {\bf 771}:
68 (2006).

\bibitem{vdw1} H. M\"uller and B. D. Serot, Phys. Rev. C {\bf 52}: 2072 (1995).

\bibitem{chomaz} Ph. Chomaz, C. Colonna, and J. Randrup, Phys. Rep. {\bf 389}: 263 
(2004). 
  
\bibitem{vdw5} J. B. Silva, O. Louren\c{c}o, A. Delfino, J. S. S\'a Martins, M. Dutra, 
Phys. Lett. B {\bf 664} 246, (2008).

\bibitem{vdw2} V. Vovchenko, D. V. Anchishkin, and M. I. Gorenstein, Phys. Rev. C {\bf 
91}: 064314 (2015).

\bibitem{vdw3} V. Vovchenko, D. V. Anchishkin, M. I. Gorenstein, and R. V. Poberezhnyuk, 
Phys. Rev. C {\bf 92}: 054901 (2015).

\bibitem{vdw6} V. Vovchenko, Phys. Rev. C {\bf 96}: 015206 (2017).

\bibitem{baldo} M. Baldo, G. F. Burgio, Prog. Part. Nucl. Phys. {\bf 91}: 203 (2016).
  
\bibitem{Piekarewicz_2001}
C. J. Horowitz and J. Piekarewicz, Phys. Rev. Lett. {\bf 86}: 5647 (2001).

\bibitem{skin_2007} S. S. Avancini, J. R. Marinelli, D. P. Menezes, M.~M.~W.~Moraes and 
C. 
Provid\^encia, Phys. Rev. C {\bf 75}: 055805 (2007).

\bibitem {Lopes_2014} L. L. Lopes and D. P. Menezes, Braz. Jour. Phys. {\bf 44}: 774 
(2014). 

\bibitem{Cavagnoli_2011} R. Cavagnoli, D. P. Menezes and C. Provid\^encia, Phys. 
Rev. C {\bf 84}: 065810 (2011).

\bibitem{Prafulla_2012} P.~K.~Panda, A.~M.~S. Santos, D. P. Menezes and C. Provid\^encia, 
Phys. Rev. C {\bf 85}:
055802 (2012).

\bibitem{EPJA_2014} C. Provid\^encia {\it et. al.}: Eur. Phys. J. A {\bf 50}: 44 (2014).

\bibitem {pais_2016} H. Pais, A. Sulaksono, B. K. Agrawal, and C. Provid\^encia, Phys. 
Rev. C {\bf 93}: 045802 (2016).

\bibitem{pasta_2009} S. S. Avancini, L. Brito, J. R. Marinelli, D. P. Menezes, 
M.~M.~W.~de~Moraes, C. Provid\^encia and A.~M. Santos, Phys. Rev. C {\bf 79}: 035804 
(2009).

\bibitem{Steiner_2009} M. B. Tsang, Yingxun Zhang, P. Danielewicz, M. Famiano, Zhuxia Li, 
W. G. Lynch, and A. W. Steiner, Phys. Rev. Lett. {\bf 102}: 122701 (2009).

\bibitem{Steiner_2010} Andrew W. Steiner, James M. Lattimer, and Edward F. Brown, 
Astrophys J. {\bf 722}: 33 (2010).

\bibitem{gamma025} Or Hen, Bao-An Li, Wen-Jun Guo, L.B. Weinstein and Eliezer Piasetzky, 
Phys. Rev. C {\bf 91}: 025803 (2015).

\bibitem{gamma072} P. Russotto et al, Phys. Rev. C {\bf 94}: 034608 (2016).

\bibitem{ppnp} J. Meng, H. Toki, S.~G. Zhou, S.~Q. Zhang, W.~H. Long, L.~S. Geng, Prog. 
Part. Nucl. Phys. {\bf 57}: 470 (2006). 

\bibitem{bka} B. K. Agrawal, Phys. Rev. C {\bf 81}: 034323 (2010).

\bibitem{bsr} S. K. Dhiman, R. Kumar, and B. K. Agrawal, Phys. Rev. C {\bf 76}: 045801 
(2007).

\bibitem{PRC85-024302} B.-J. Cai, L.-W. Chen, Phys. Rev. C {\bf 85}: 024302
(2012).

\bibitem{fsugold} B. G. Todd-Rutel and J. Piekarewicz, Phys. Rev. Lett. {\bf 95}: 122501 
(2005).

\bibitem{fsugold4} J. Piekarewicz and S. P. Weppner, Nucl. Phys. A {\bf 778}: 10 (2006).

\bibitem{fsugz} R. Kumar, B. K. Agrawal, and S. K. Dhiman, Phys. Rev. C {\bf 74}: 034323 
(2006).

\bibitem{g2*} A. Sulaksono and T. Mart, Phys. Rev. C {\bf 74}: 045806 (2006).

\bibitem{PRC82-055803} F. J. Fattoyev, C. J. Horowitz, J. Piekarewicz, and G.
Shen, Phys. Rev. C {\bf 82}: 055803 (2010).

\bibitem{z271} C. J. Horowitz and J. Piekarewicz, Phys. Rev. C {\bf 66}: 055803 (2002).

\bibitem{ddf} T. Kl\"ahn, {\it et al.}: Phys. Rev. C {\bf 74}: 035802 (2006).

\bibitem{tw99} S. Typel and H. H. Wolter, Nucl. Phys. A {\bf 656}: 331 (1999).

\bibitem{ddhd} T. Gaitanos, M. Di Toro, S. Typel, V. Baran, C. Fuchs, V. Greco, and H. H. 
Wolter, Nucl. Phys. A {\bf 732}: 24 (2004).

\bibitem{ddmed} X. Roca-Maza, X. Vi\~nas, M. Centelles, P. Ring, and P. Schuck, Phys. 
Rev. C {\bf 84}: 054309 (2011).

\bibitem{fa3} J. J. Rusnak and R. J. Furnstahl, Nucl. Phys. A {\bf 627}: 495 (1997).

\bibitem{pc1} B. A. Nikolaus, T. Hoch, and D. G. Madland, Phys. Rev. C {\bf 46}: 1757 
(1992). 

\bibitem{pc2} D. G Madland, T. J B\"urvenich, J. A Maruhn, P.-G Reinhard, Nucl. Phys. A 
{\bf 741}: 52 (2004).

\bibitem{pc3} O. Louren\c{c}o, M. Dutra, A. Delfino, and R. L. P. G. Amaral, Int. 
Jour. Mod. Phys. E, {\bf 16}: 3037 (2007). 

\bibitem{pc4} T. Niksic, D. Vretenar, and P. Ring, Prog. in Part. and Nucl. Phys. {\bf 
66}: 519 (2011).

\bibitem{Lattimer_2013} J. M. Lattimer and Y. Lim, Astrophys. J. {\bf 51}: 771 (2013).

\bibitem{NScrust_book} {\it Neutron Star Crust}: editors Carlos
  Bertulani and Jorge Pieckarewicz, Nova Science Publishers, 2012, New York.

\bibitem{bianca} B. M. Santos, M. Dutra, O. Louren\c co, and A. Delfino, Phys. Rev. C 
{\bf 90}: 035203 (2014); {\bf 92}: 015210 (2015).

\bibitem{src1} R. Subedi {\it et. al.}: Science {\bf 320}: 1476 (2008).

\bibitem{src2} O. Hen {\it et al.}: Science {\bf 346}: 614 (2014).

\bibitem{src3} O. Hen, G. A. Miller, E. Piasetzky, and L. B. Weinstein, to apear in 
Rev. Mod. Phys (2017).

\bibitem{brandt} S. Brandt, {\it Data Analysis: Statistical and Computational Methods for 
Scientists and Engineers}: 4th ed. (Springer, New York, 2014).

\bibitem{constanca} N. Alam, B. K. Agrawal, M. Fortin, H. Pais, C. Provid\^encia, Ad. R. 
Raduta, and A. Sulaksono, Phys. Rev. C {\bf 94}: 052801(R) (2016).

\bibitem{bps} G.~Baym, C.~Pethick, P.~Sutherland, Astrophys. J. \textbf{170}: 299 (1971).

\bibitem{tov} J.~R.~Oppenheimer, G.~M.~Volkoff, Phys. Rev. \textbf{33}: 374 (1939).

\bibitem{details} N.~K. ~Glendenning, {\it Compact Stars}: 2nd ed. (Springer, New 
York, 2000); P. Haensel, A. Y. Potekhin, and D. G. Yakovlev, {\it Neutron Stars,
Equation of State and Structure} (Springer, New York, 2006).

\bibitem{Bao_book} Bao-An Li, W. Udo Schr\"oder (Eds.), {\it Isospin
    Physics in Heavy-Ion Collisions at Intermediate  Energies}: Nova
Science Publishers, Huntington, NY, 2001.

\bibitem{Bao_2004} Bao-An Li, Champak B. Das, Subal Das Gupta, Charles
  Gale, Nuc. Phys. A {\bf 735}: 563 (2004).
 
\end{thebibliography}
\end{document}